\documentclass[12pt]{article}
\usepackage{exscale}
\usepackage{amsmath}
\usepackage{amsfonts}
\usepackage{times}
\usepackage{amsthm}
\usepackage{epsfig}
 \newcommand{\DATUM}{24-Jul-2006}            
\def\bbbone{{\mathchoice {\rm 1\mskip-4mu l} {\rm 1\mskip-4mu l}
{\rm 1\mskip-4.5mu l} {\rm 1\mskip-5mu l}}}
\newcommand{\one}{\bbbone}

\newcommand{\comma}{\: ,}              
\newcommand{\period}{\: .}             
\newcommand{\Proof}{\noindent\emph{Proof. }}              
\newcommand{\eps}{{\varepsilon}}        
\newcommand{\vphi}{{\varphi}}           
\newcommand{\om}{\omega} 
\newcommand{\Om}{\Omega}
\newcommand{\pOm}{{\partial\Omega}}
\newcommand{\deltaE}{{\delta E}}
\newcommand{\la}{\langle} 
\newcommand{\ra}{\rangle}
\newcommand{\ol}{\overline} 
\newcommand{\ul}{\underline}
\newcommand{\ua}{{\uparrow}} 
\newcommand{\da}{{\downarrow}} 
\newcommand{\uda}{{\uparrow, \downarrow}} 
\newcommand{\cE}{\mathcal{E}}
\newcommand{\cF}{\mathcal{F}}

\newcommand{\RR}{\mathbb{R}}            
\newcommand{\CC}{\mathbb{C}}            
\newcommand{\TT}{\mathbb{T}}
\newcommand{\ZZ}{\mathbb{Z}}
\newcommand{\hDelta}{\widehat{\Delta}}

\newcommand{\tcE}{\widetilde{\cE}}

\newcommand{\tP}{\widetilde{P}}
\newcommand{\tW}{\widetilde{W}}

\newcommand{\tmu}{{\tilde{\mu}}}

\newcommand{\trho}{\widetilde{\rho}}

\newcommand{\tDelta}{\widetilde{\Delta}}

\newcommand{\Ran}{\mathop{\mathrm{Ran}}}     
\newcommand{\dist}{{\mathrm{dist}}}          
\newcommand{\Tr}{\mathrm{Tr}}

\newcommand{\cre}{c^*}
\newcommand{\ann}{c^{\;}}
\newcommand{\vac}{|0\ra}
\newcommand{\gs}{\mathrm{gs}}
\newcommand{\hfz}{{(\mathrm{hfz})}}

\newcommand{\re}{\mathrm{Re}}
\renewcommand{\thesection}
{\arabic{section}}                     
\renewcommand{\theequation}
{\thesection.\arabic{equation}}        
\newcommand{\secct}[1]{\section{#1}
\setcounter{equation}{0}}              

\newtheorem{theorem}{Theorem}[section]         
\newtheorem{lemma}[theorem]{Lemma}             
\theoremstyle{plain}
\begin{document}
\bibliographystyle{unsrt}
%

\title{Ferromagnetism of the Hubbard Model at Strong Coupling in the
  Hartree-Fock Approximation} 
\author{
  Volker Bach \\
  \small{Institut f\"ur Mathematik; Universit\"at Mainz;} \\[-1ex]
  \small{D-55099 Mainz; Germany (vbach@mathematik.uni-mainz.de)} \and
  Elliott H.~Lieb \\
  \small{Departments of Mathematics and Physics; Jadwin Hall;
    Princeton University;} \\[-1ex]
  \small{P.O.~Box 709; Princeton NJ 08544; USA
    (lieb@math.princeton.edu)} \and
  Marcos V.~Travaglia\\
  \small{Institut f\"ur Mathematik; Universit\"at Mainz;} \\[-1ex]
  \small{D-55099 Mainz; Germany (marcos@mathematik.uni-mainz.de)} }
    
\date{\DATUM}
\maketitle
\begin{abstract}
  As a contribution to the study of Hartree-Fock theory we prove
  rigorously that the Hartree-Fock approximation to the ground state
  of the $d$-dimensional Hubbard model leads to saturated
  ferromagnetism when the particle density (more precisely, the
  chemical potential $\mu$) is small and the coupling constant $U$ is
  large, but finite. This ferromagnetism contradicts the known
  fact that there is no magnetization at low density, for any $U$, and
  thus shows that HF theory is wrong in this case. As in the usual
  Hartree-Fock theory we restrict attention to Slater determinants
  that are eigenvectors of the z-component of the total spin,
  $\mathbb{S}_z = \sum_x n_{x,\uparrow} - n_{x,\downarrow}$, and we
  find that the choice $2\mathbb{S}_z=N=$ particle number gives the
  lowest energy at fixed $0 < \mu < 4d$. 
\\[1ex]
\ul{Keywords:} Hubbard model, Ferromagnetism, Hartree-Fock Theory
\\[1ex]
\ul{AMS Math.~Subj.~Class.~2000:} 82D40 (Magnetic materials)
\end{abstract}

\newpage

\secct{Introduction} \label{sec-I}
%
The (one-band) Hubbard model has become a standard model for
correlated electrons in condensed matter physics since it is, perhaps,
the simplest possible model of itinerant interacting electrons. In
spite of its simplicity, its zero temperature phase diagram is rich
with different magnetic phases such as paramagnetic, ferromagnetic,
and antiferromagnetic phases,
depending on the details of the hopping amplitudes, the (relative)
coupling constant $U/t$ and the filling parameter $\nu =
N/(2|\Lambda|)$.

As the Hubbard model is a many-body fermion model, the computation of
its ground state for large lattices is a difficult, if not impossible,
task, except in one-dimension \cite{LiebWu1967,LiebWu2003}. Thus various
schemes have been developed during the past decades to derive an
approximate ground state and then to study its magnetic phase diagram.

In the present paper, we consider the Hartree-Fock approximation of
the (repulsive, one-band, nearest-neighbor-hopping) Hubbard model with
the intention of studying the validity of the Hartree-Fock
approximation. We require the Slater determinants entering the
Hartree-Fock energy functional to be eigenfunctions of the operator
$\mathbb{S}_z := \sum_{x \in \Lambda} \{ n_{x,\ua}-n_{x,\da} \}$ of
total spin in the $z$-direction, and for this reason we refer to the
model as the {\it HFz approximation}. Our requirement means that each
orbital has the form $\vphi(x) \otimes | \ua \ra$ or $\vphi(x) \otimes
|\da \ra$. This is a restriction in the sense that general orbitals
are of the form $\vphi(x, \sigma)$, in which the spin direction
depends on position. No other restriction is imposed on the
variational states; in particular, no assumption about translation
invariance is made a priori. For the HFz model, at small chemical
potential and for sufficiently strong repulsion, we give a
mathematical proof of \emph{saturated ferromagnetism} in the
Hartree-Fock ground state. That is, the HF ground state has maximal
total spin and maximal ferromagnetic long-range spatial order. The
smallness of the chemical potential and the large strength of the
repulsion also insure that the HF ground state density is strictly
below half-filling.
%

Before we come to a detailed description of our result and its proof,
we discuss it in comparison to other works. 

The appearance of ferromagnetic behaviour has been anticipated in many
studies of the Hubbard model and approximations thereof. Among these
are (restricted) Hartree-Fock approximations \cite{Penn1966}, DMFT
models in the limit of infinite spatial dimension
\cite{VanDongen1991,VanDongen1994,ObermeierPruschkeKeller1997,
  WahleBluemerSchlipfHeldVollhardt1997}, exact diagonalizations on
small lattices \cite{PastorHirschMuehlschlegel1996}, variational
calculations \cite{HanischUhrigMueller-Hartmann1997} and studies at
low filling \cite{Mueller-Hartmann1995}. These studies support the
conjecture that, for large coupling $U/t \gg 1$ and away from
half-filling, $\nu \neq 1/2$, the ground state of the Hubbard model is
ferromagnetic. Ferromagnetism has been established for the (full)
Hubbard model in case the dispersion relation leads to a very high
density of states around the Fermi energy
\cite{Tasaki1992,Mielke1993,MielkeTasaki1993} and in case of
next-nearest-neigbor hopping \cite{Tasaki1995,Tasaki2003}.

As said before, the main purpose of the present paper is to prove
ferromagnetic behaviour with mathematical rigor. None of the papers
\cite{VanDongen1991,VanDongen1994,ObermeierPruschkeKeller1997,
  WahleBluemerSchlipfHeldVollhardt1997,PastorHirschMuehlschlegel1996}
cited above match the standards of a mathematical proof: The orbitals
in the Hartree-Fock approximation are a priori assumed to be composed
of only few Fourier modes; the error terms when taking the limit of
infinite spatial dimension in DMFT are not under control; exact
diagonalizations are restricted to very small lattices and the
implication of these to the thermodynamic limit remains unclear. The
work by Mielke and Tasaki
\cite{Tasaki1992,Mielke1993,MielkeTasaki1993} is mathematically
rigrous, but the assumptions made therein about the lattice structure
are rather special. On the other hand, by adding next-nearest-neigbor
hopping (two-band Hubbard model), Tasaki \cite{Tasaki1995,Tasaki2003}
has found a Hubbard model that displays ferromagnetism in all
dimensions. Tasaki also reviews rigorous results on ferromagnetism 
in the Hubbard model in \cite{Tasaki1998}.

While the prediction of ferromagnetism in the Hubbard model and
approximations thereof is supported by the above studies, we also know
that HF theory predicts {\it anti-}ferro\-mag\-ne\-tism (in the sense
that the total spin is zero) at higher densities, notably at
half-filling \cite{BachLiebSolovej1994}.  Furthermore, our proof shows
saturated ferromagnetism at low density and sufficiently large
coupling in HF theory, even in one-dimension, but the actual ground
state {\it always} has spin zero in one-dimension as long as there is
only nearest-neighbor hopping (see \cite{LiebMattis1962a}).

Even more seriously, our conclusion is opposite to what {\it actually}
occurs in the Hubbard model. Namely, at very low density (and
independent of the value of $U > 0$), there is no magnetization in the
ground state of this model. In the ground state $\mathbb{S}_z$ is
close to zero and converges to zero, as the particle density tends to
zero. This has been pointed out in
\cite{PieriDaulBaeriswylDzierzawaFazekas1996,Tasaki1998}, based on
arguments similar to the following transcription to lattice systems
of the recent work \cite{LiebSeiringerSolovej2005}.

In this paper \cite{LiebSeiringerSolovej2005} it was shown that
fermions in the 3-dimensional continuum $\RR^3$ (instead of the
lattice $\ZZ^3$), and with a repulsive two-body potential, have a
ground state energy density, $e$, given by
\begin{equation} \label{eq-I.0,1}
e(\rho_\uparrow , \rho_\downarrow ) = \frac{\hbar^2}{2m}
\frac{3}{5}(6\pi^2)^{2/3}
\left(\rho_\uparrow^{5/3} + \rho_\downarrow^{5/3}\right)
+ \frac{\hbar^2}{2m}
 8\pi a \rho_\uparrow  \rho_\downarrow
+\mathrm{higher\ order\ in\ } (\rho_\uparrow , \rho_\downarrow) \ ,
\end{equation}
where $\rho_\uparrow , \rho_\downarrow$ are the densities of the
`spin-up' and the `spin-down' fermions and $a$ is the scattering
length of the two-body potential. Because $\rho^{5/3}$ dominates
$\rho^2$ for small $\rho$, it is clear from (\ref{eq-I.0,1}) that the
minimum energy occurs approximately, if not exactly, when
$\rho_\uparrow = \rho_\downarrow = \rho/2$. This answers the questions
in \cite[problem~3]{Lieb1994}.

To show that there is vanishing net magnetization as $\rho \to 0$ one
only needs an upper bound for $e$ of the form (\ref{eq-I.0,1}). For
the Hubbard model (where the two-body potential is a positive
delta-function, or even a hard core) this can conveniently be done by
a variational wave-function of the form $\Psi = F \Psi_0$, where
$\Psi_0$ is a Slater determinant, and $F$ is the projection onto the
states with no double occupancy -- in imitation of
\cite{PieriDaulBaeriswylDzierzawaFazekas1996,Tasaki1998,
  LiebSeiringerSolovej2005}. We omit the details, but we draw
attention to the fact that $F \Psi_0$ is not a Slater determinant,
reflecting the more complex structure of correlations in the actual
ground state of the Hubbard model. The proof of an analog of
(\ref{eq-I.0,1}) with precise constants is a more complicated matter
which is now under investigation, but it is not needed for the present
discussion.

Our setting is the usual (repulsive) Hubbard model with
nearest-neighbor hopping on a $d$-dimensional cubic lattice $\Lambda$,
with periodic boundary conditions and linear size $L$, which we assume
to be an even integer.  It is defined by the second quantized
Hamiltonian
\begin{equation} \label{eq-I.1}
H_{\mu,U} \ = \ 
\sum_{x,y \in \Lambda, \: \sigma = \uda} 
(-\Delta_{x,y} - \mu \delta_{x,y}) \, \cre_{x,\sigma} \ann_{y,\sigma}
\; + \; U \sum_{x \in \Lambda} n_{x,\ua} \, n_{x,\da} \period
\end{equation}
We work at fixed chemical potential $\mu$ instead of fixed particle
number. The only slightly unusual notation is $\Delta_{x,y} = T_{x,y}
- 2d \delta_{x,y}$ for the matrix elements of the discrete Laplacian
$\Delta$ on $\Lambda$, with $T_{x,y} := \one[ |x-y|_1=1 ]$ being the
nearest-neighbor hopping matrix and $\delta_{x,y} = \one[x=y]$ the
Kronecker-Delta.

The operators $\cre_{x,\sigma}$, $\ann_{x,\sigma}$, and $n_{x, \sigma}
:= \cre_{x,\sigma} \ann_{x,\sigma}$ are the usual fermion creation,
annihilation, and number operators, respectively, at site $x \in
\Lambda$ and of spin $\sigma \in \{ \ua, \da \}$, obeying the
canonical anticommutation relations $\{ \ann_{x,\sigma} ,
\ann_{y,\tau} \} = \{ \cre_{x,\sigma} , \cre_{y,\tau} \}$ $= 0$, $\{
\ann_{x,\sigma} , \cre_{y,\tau} \} = \delta_{x,y}
\delta_{\sigma,\tau}$, and $\ann_{x,\sigma} \vac = 0$, for all
$x,y,\sigma,\tau$. Here $\vac$ is the vacuum vector in the usual Fock
space $\cF_\Lambda := \cF_f(\CC^\Lambda \otimes \CC^2)$ of
spin-$\frac{1}{2}$ fermions. The Hamiltonian $H_{\mu,U}$ depends
parametrically on the chemical potential $\mu >0$ and the coupling
constant $U >0$.

Note that the usual hopping parameter $t$ equals $1$ here and that the
discrete Laplacian $\Delta$ differs from the usual hopping matrix by
the inclusion of the diagonal term, i.e., $2d$ times the identity
matrix. This difference amounts to a convenient redefinition of the
chemical potential $\mu$, so that $\mu=0$ corresponds precisely to zero
filling since the hopping matrix $-\Delta \geq 0$ is a positive
semi-definite matrix. Moreover, the boundedness $0 < \mu < 4d$
of $\mu$ together with the assumption that $U \gg 4d$ insures that the
corresponding electron density in the HF ground state is always at
{\it low filling}, i.e., strictly below half-filling, $0 \leq \rho <
1$.

Our definition of $\mu$ is convenient because in this
paper, we are concerned with the Hubbard model at {\it low filling}, 
and Our assumption
of a bounded chemical potential $0 \leq \mu \leq 2d$

Apart from this,
everything is standard.

The Hamiltonian $H_{\mu,U}$ is a linear operator on the Fock space and
the ground state energy $E_{\mu,U}^{(\gs)}$ is its smallest eigenvalue,
\begin{equation} \label{eq-I.2}
E_{\mu,U}^{(\gs)} \ := \ \min\big\{ \la \Psi | \: H \, \Psi \ra \ \big| \
\Psi \in \cF_\Lambda, \; \|\Psi\| =1 \big\} \period
\end{equation}
As the dimension $\dim(\cF_\Lambda) = 2^{\dim(\CC^\Lambda \otimes
  \CC^2)} = 4^{(L^d)} < \infty$ is finite, the determination of
$E_{\mu,U}^{(\gs)}$ amounts to diagonalizing the finite-dimensional,
selfadjoint matrix $H_{\mu,U}$.  The fast growth of this dimension
with the number $L^d$ of points in the lattice $\Lambda$, however,
allows for an explicit diagonalization of $H_{\mu,U}$ by a modern
computer only up to $L=4$, in three spatial dimensions, $d=3$.

The Hartree-Fock (HF) approximation is an important method to reduce
the high-dimensional many-particle problem given by the
diagonalization of $H_{\mu,U}$ to a low-dimensional, but nonlinear
variational problem.  It is defined by restricting the minimization in
(\ref{eq-I.2}) to Slater determinants $\vphi_1 \wedge \cdots \wedge
\vphi_N$, where $\{ \vphi_i \}_{i=1}^N \subseteq \CC^\Lambda \otimes
\CC^2$ is an orthonormal family of $N $ one-electron wave functions.
The HF approximation to the Hubbard model was analyzed in
\cite{BachLiebSolovej1994} in the special situation when the number
of electrons equals the number of lattice sites, $N = |\Lambda|$,
which is usually referred to as \emph{half-filling}.

Note that a priori no other condition but orthonormality is imposed on
the orbitals $\{ \vphi_i \}_{i=1}^N$ in the Slater determinants varied
over in Hartree-Fock theory. This is sometimes stressed by calling it
the \emph{unrestricted Hartree-Fock theory}. Let us temporarily
consider a general many-body Hamiltonian $H$ which commutes with a
certain symmetry operator $S$, i.e., $[H,S] = 0$. It is important to
note that in this case, the HF ground state $\Phi_{hf}$, i.e., the
Slater determinant which minimizes the energy $\la \Phi_{hf} | H \,
\Phi_{hf} \ra$, is not necessarily an eigenstate of $S$. Phrased
differently, unrestricted Hartree-Fock theory may (depending on the
model) break the symmetry $S$. The following are examples that occur
in physically relevant situations: unrestricted HF ground states of
atoms are, in general, not eigenfunctions of the angular momentum
operator (because in unrestricted HF theory, all shells are filled
\cite{BachLiebLossSolovej1994}) - even though the atomic Hamiltonian
is rotationally invariant; the ground state in the BCS theory of
superconductors (which is a variant of HF theory) is not an
eigenfunction of the number operator - even though the BCS Hamiltonian
preserves the particle number; a HF ground state for the Hubbard model
with non-zero spin breaks the invariance of the Hubbard Hamiltonian
under global spin rotations; charge density waves (CDW) and spin
density waves (SDW) of the Hubbard model are translation invariant
only by translation of an \emph{even} number of lattice sites,
breaking the (full) translation symmetry the Hubbard Hamiltonian
$H_{\mu,U}$ posesses. As it is impossible to predict a priori whether
a symmetry of the Hamiltonian is preserved or not, we call all
variations of $\la \Phi | H \, \Phi \ra$ over Slater determinants
$\Phi$ which fulfill an additional constraint \emph{restricted
  Hartree-Fock theory}.

In this paper, we  consider a restricted Hartree-Fock theory,
which we term the \emph{HFz approximation}. The further restriction
imposed is that we minimize in (\ref{eq-I.2}) only over Slater
determinants $\Phi$ that are eigenfunctions of the operator
$\mathbb{S}_z := \sum_{x \in \Lambda} \{ n_{x,\ua}-n_{x,\da} \}$ of
total spin in the $z$-direction. One could rephrase our condition by
saying that we do not allow for spiral spin density waves (SSDW; see,
e.g., \cite{Penn1966}) in (\ref{eq-I.2}). Once again, it is customary
to employ this restriction in HF calculations without explicitly
drawing attention to the fact that this is a restriction. (In
\cite{BachLiebSolovej1994} mentioned above, however, we dealt with
truly unrestricted HF theory.)

More concretely, our HF wave functions have the form
\begin{equation} \label{eq-I.3}
\Phi \ = \ \prod_{i=1}^{N_\ua} c_\ua^*(f_i)  
           \prod_{j=1}^{N_\da} c_\da^*(g_i) \: \vac \comma
\end{equation}
where $c_\uda^*(f) = \sum_{x \in \Lambda} f(x) \, c_{x,\uda}^*$, the integers
$N_\uda$ are the particle numbers, and where the $f_i $ and $g_i $ 
are two families of orthonormal wave functions on the lattice $\Lambda$, i.e., 
$\la f_i | f_j \ra = \la g_i | g_j \ra = \delta_{i,j}$, with
$\la f | g \ra := \sum_{x \in \Lambda} \ol{f(x)} g(x)$ denoting the 
usual hermitian scalar product for such functions. 

It is convenient to rephrase the HFz approximation in terms of
one-particle density matrices, i.e., complex, self-adjoint $\Lambda
\times \Lambda$ matrices whose eigenvalues lie between $0$ and $1$.
To this end, we denote
\begin{equation} \label{eq-I.3,1}
K_\mu \ := \ -\Delta - \mu 
\end{equation}
and observe that
\begin{eqnarray} \label{eq-I.4}
\la \Phi | \: H \, \Phi \ra 
& = &
\sum_{i=1}^{N_\ua} \la f_i | \: K_\mu \, f_i \ra 
\; + \;
\sum_{j=1}^{N_\da} \la g_j | \: K_\mu \, g_j \ra 
\nonumber \\[1mm]
& & 
\; + \; 
U \, \sum_{x \in \Lambda} 
\Big( \sum_{i=1}^{N_\ua } |f_i(x)|^2 \Big) \, 
\Big( \sum_{j=1}^{N_\da} |g_j(x)|^2 \Big) \period
\end{eqnarray}
Introducing the one-particle density matrices 
$\gamma_\uda $ corresponding to $\Phi$ by
\begin{equation} \label{eq-4}
\gamma_\ua \ := \ \sum_{i=1}^{N_\ua} |f_i\ra \la f_i| 
\hspace{6mm} \mbox{and} \hspace{6mm}
\gamma_\da \ := \ \sum_{i=1}^{N_\da} |g_i\ra \la g_i| \comma
\end{equation}
we observe that $\gamma_\uda = \gamma_\uda^* =  \gamma_\uda^2$ are
orthogonal projections of dimension  $N_\uda$ and that
the energy expectation value of the Slater determinant $\Phi$ is given by 
$\la \Phi | \: H \, \Phi \ra = \cE_{\mu,U}^\hfz (\gamma_\ua, \gamma_\da)$, 
where
\begin{equation} \label{eq-I.5}
\cE_{\mu,U}^\hfz (\gamma_\ua, \gamma_\da)
\ := \ 
\Tr\big\{ K_\mu \, (\gamma_\ua + \gamma_\da) \big\}
\; + \;
U \, \sum_{x \in \Lambda} \rho_\ua(x) \, \rho_\da(x) \comma
\end{equation}
and the diagonal matrix elements $\rho_\uda(x) := (\gamma_\uda)_{x,x}$
of $\gamma_\uda$ are the one-particle densities of the electron with
spin up (``$\ua$'') and spin down (``$\da$''), respectively. 

The symbol ``$\Tr$'' denotes the usual trace 
$\Tr\{A\} = \sum_{x \in \Lambda} A_{x,x}$ 
of a complex $\Lambda \times \Lambda$ matrix 
$A = (A_{x,y})_{x,y \in \Lambda}$ with $A_{x,y} \in \CC$. 
That is, ``$\Tr$'' is the trace over the states in $\CC^\Lambda$ of a single
spinless particle on the lattice $\Lambda$. It does not include spin
states, and it is not the trace over states in Fock space.

Let us note that the particle numbers $N_\uda$ are not determined {\it
  ab initio}. We are in the grand canonical ensemble, so they are
determined by the condition that the total energy (\ref{eq-I.5}) is
minimized.

These observations motivate us to define the \emph{HFz energy} by the
following variational principle over projections:
\begin{equation} \label{eq-I.6}
E_{\mu,U}^\hfz 
\ := \ 
\min\big\{ \cE_{\mu,U}^\hfz (\gamma_\ua, \gamma_\da) \; \big|
\ \gamma_\uda = \gamma_\uda^* =  \gamma_\uda^2 \big\} \period
\end{equation}
The two sets of orthogonal projections on $\CC^\Lambda$ over which we
minimize in (\ref{eq-I.6}) is not really well-suited for a variational
analysis.  In particular, they are not convex. An observation in
\cite{Lieb1981a}, however, states that, because $U\geq 0$, we will
obtain the same value for the minimum if we vary over the larger set
of {\it all} one-particle density matrices, $0 \leq \gamma_\uda \leq
1$, not only over projections. (Recall that a density matrix is a
hermitean $\Lambda \times \Lambda$ matrix $\gamma$ whose eigenvalues
lie between 0 and 1, i.e., $0 \leq \gamma \leq 1$, as a matrix
inequality.) Our extended $E_{\mu,U}^\hfz $ is then
\begin{equation} \label{eq-I.7}
E_{\mu,U}^\hfz 
\ = \ 
\min\big\{ \cE_{\mu,U}^\hfz (\gamma_\ua, \gamma_\da) \; \big|
\ 0 \leq \gamma_\uda \leq 1 \big\} \period
\end{equation}
The evaluation of $E_{\mu,U}^\hfz$ and the determination of those
pairs $(\gamma_\ua, \gamma_\da)$ of one-particle density matrices that
minimize $\cE_{\mu,U}^\hfz$ is the objective of this paper. Our main
result is that, for any $0 < \mu < 4d$, the minimal value of
$\cE_{\mu,U}^\hfz$ is attained for the saturated ferromagnet, provided
$U < \infty$ is sufficiently large.
\begin{theorem}[Ferromagnetism] \label{thm-I.1} 
For any $0 < \mu < 4d$, there is a finite length $L_\#(\mu) $ and a finite
coupling constant $U_\#(\mu) \geq 0 $, such that, for all even 
$L \geq L_\#(\mu)$ and all $U \geq U_\#(\mu)$, the minimal HFz energy
is given by the sum of the negative eigenvalues of $-\Delta - \mu$,
\begin{equation} \label{eq-I.9}
E_{\mu,U}^\hfz \ = \ \Tr\big\{ [-\Delta - \mu]_- \big\} \period
\end{equation}
If $\mu$ is not an eigenvalue of $-\Delta$ and if 
$(\gamma_\ua, \gamma_\da)$ is a minimizer of the HFz functional,  
i.e., $0 \leq \gamma_\uda \leq 1$, and 
$\cE_{\mu,U}^\hfz(\gamma_\ua, \gamma_\da) = E_{\mu,U}^\hfz$, then
\begin{eqnarray} \label{eq-I.10a}
\mbox{either}
& \gamma_\ua \: = \: \one[-\Delta < \mu] \comma 
& \gamma_\da \: = \: 0
\\ \label{eq-I.10b}
\mbox{or}
& \gamma_\ua \: = \: 0 \comma
& \gamma_\da \: = \: \one[-\Delta < \mu] \comma 
\end{eqnarray}
where $\one[-\Delta < \mu]$ is the spectral projection of $-\Delta$ onto 
$(-\infty , \mu)$.
\end{theorem}
[With reference to Eq. (\ref{eq-I.9}) and elsewhere, note that in our
notation, $[X]_- = \min\{X,\, 0 \}$ is negative, whereas elsewhere one
often defines $[X]_-$ to be positive, i.e., $[X]_- := \max\{-X,\, 0\}$. 
If $X$ is a self adjoint operator then $[X]_-$ denotes the negative
part of $X$ and $\Tr [X]_-$ is the sum of the negative eigenvalues of
$X$.]

Theorem~\ref{thm-I.1} is not really as complicated as it looks. It is
stated in terms of a length $L_\#$ and coupling constant $U_\#$ in
order to make it clear that the state of saturated ferromagnetism is
obtained not only asymptotically in the thermodynamic limit and
asymptotically as $U \to \infty$, but it holds for all systems with
large, finite interaction and sufficiently large size.

Theorem~\ref{thm-I.1} states that, for \emph{any} value of the
chemical potential $\mu \in (0,4d)$, the HFz variational 
principle yields a ferromagnetic minimizer, provided $U$ and $L$
are chosen sufficiently large (but still finite). A similar statement
was proved in \cite[Theorem~4.7]{BachLiebSolovej1994} for $U = \infty$
(which amounts to requiring $\la \Phi | n_{x,\ua} n_{x,\da} \Phi \ra = 0$,
on every lattice site $x \in \Lambda$).

At first sight, Theorem~\ref{thm-I.1} seems to contradict another fact
proved in \cite{BachLiebSolovej1994} that the HF minimizer is
antiferromagnetic at half-filling. But as the definition of
the chemical potential $\mu$ in present paper differs from
its definition in \cite{BachLiebSolovej1994} by $2d+U$, the
parameter range of the present paper and of \cite{BachLiebSolovej1994}
never overlap and, hence, there is no contradiction.

As just mentioned, the minimal HF energy and the minimal HFz energy
agree in the half-filling case, as shown in \cite{BachLiebSolovej1994}.
We conjecture that this is also the case for the range 
of the chemical potential $\mu \in (0,4d)$ and sufficiently
large $U$, but we do not know how to prove this conjecture. This
is a topic for future research.

From Theorem~\ref{thm-I.1} we conclude that at small filling there is
a phase transition (within the context of HFz theory) from
paramagnetism for small $U$ to saturated ferromagnetism for large $U$.
This follows from continuity and the fact that when $U=0$ we can find
the ground state explicitly and, as is well known, it has $S=0$ and is
obtained from filling up the Fermi sea for both $\ua$ and $\da$
states.

If $0 < \mu \leq \frac{1}{2}$ then we can estimate $L_\#(\mu) $ and
$U_\#(\mu) $ in Theorem~\ref{thm-I.1} more explicitly. For the precise
formulation of these estimates, we introduce the following constants,
\begin{eqnarray} \label{eq-I.8,1}
L_*(\mu) 
& := & 
2 \, M_*(\mu) 
\ \; := \ \; 
24  \, (4d)^2 \, \mu^{-2} \comma
\\ \label{eq-I.8,2}
\kappa(\mu)
& := &
\frac{\mu^d}{4^{2d+1} \, e^d \, d^d} \;
\Big[1 + 2 \, \ln(2) \, (d^{-1} + 1) + \ln\big(4d \mu^{-1} \big) \Big]^{-2d} 
\\ \label{eq-I.8,3}
\alpha_*(\mu)
& := &
\frac{ |S^{d-1}| \: \mu^{(2+d)/2} }{2^{1+d/2} \, (2\pi)^d \, (4d)^5}
\\ \label{eq-I.8,4}
\delta_*(\mu, \alpha) 
& := &
\min\Big\{ \frac{\alpha^2}{(12 d)^2} \comma \ \ 
\frac{\alpha}{3\mu\, [4M_*(\mu)+1]^d} \comma \ \ 
\frac{\kappa(\mu)}{2} \Big\}
\\ \label{eq-I.8,5}
U_*(\mu, \alpha) 
& := & 
\max\Big\{ \frac{2\mu}{\delta_*(\mu, \alpha)} \comma \ \ 
\frac{24 d^2}{\alpha \, \delta_*(\mu, \alpha)} \Big\} \comma
\end{eqnarray}
where $|S^{d-1}| = 2\pi^{d/2}/\Gamma(d/2)$
is the measure of the unit sphere  in $\RR^d$. 
\begin{theorem} \label{thm-I.1a} 
For any $0 < \mu \leq \frac{1}{2}$, Theorem~\ref{thm-I.1}
holds true with $L_\#(\mu) := L_*(\mu)$ and 
$U_\#(\mu) := U_*(\mu, \alpha_*(\mu))$, as defined in (\ref{eq-I.8,1}),
(\ref{eq-I.8,3}), and (\ref{eq-I.8,5}).
\end{theorem}
The explicit form of $L_*(\mu)$, $\alpha_*(\mu)$, and $U_*(\mu,
\alpha_*(\mu))$, for a given $0 < \mu \leq \frac{1}{2}$, in
Theorem~\ref{thm-I.1a} allows us to estimate the actual minimal size
of $L$ and $U$ that guarantees saturated ferromagnetism. The
distinction between $\mu \leq 1/2$ and $\mu > 1/2$ is not a
fundamental one. It is an artifact of the use in Lemma \ref{lem-IId.1}
of refs.~\cite{FreericksLiebUeltschi2002} and \cite{Goldbaum2003},
whose methods favored this technical distinction.

\vspace{9mm}

\noindent \textbf{Acknowledgements:} The authors are grateful to
Alessandro Giuliani for very helpful discussions and comments about an
earlier version of this paper. They also thank Manfred Salmhofer,
J\"urg Fr\"ohlich, and Daniel Ueltschi for useful discussions. MT
thanks the german student exchange service DAAD for a generous
stipend, which supported two thirds of his graduate studies. VB and MT
gratefully acknowledge financial support from grant
no.~HPRN-CT-2002-00277 of the European Union and grant no.~Ba~1477/3-3
of the Deutsche Forschungsgemeinschaft. EL gratefully acknowledges
support from the Alexander von Humboldt Foundation of a fellowship,
the U.S. National Science Foundation, grant no.~PHY-0133984, and the
hospitality of the Mathematics Departments of the University of Mainz
and the Technical University of Berlin. The authors appreciate the
careful and helpful work of a referee.

\newpage

\secct{Proofs of Theorems~\ref{thm-I.1} and \ref{thm-I.1a}}
\label{sec-II}
%
This section contains the proofs of our main results,
Theorems~\ref{thm-I.1} and \ref{thm-I.1a}, with the aid of several
lemmas which will be proved later in Section~\ref{sec-III}. 
Here is a brief outline of the strategy of the proof.

\vspace*{1ex} \noindent $\bullet$\ We first reduce the minimization of
$\cE_{\mu,U}^\hfz(\gamma_\ua, \gamma_\da)$ in (\ref{eq-I.7}) over {\it
  two} one-particle density matrices $\gamma_\ua$ and $\gamma_\da$ to the
minimization of an effective energy functional
$\tcE_{\mu,U}^\hfz(\gamma)$ which depends only {\it one} one-particle
density matrix $\gamma$. It is given as a sum of two terms,
$\tcE_{\mu,U}^\hfz(\gamma) = \Tr\{ K_\mu \gamma \} + \Tr\{ [K_\mu + U
\rho]_- \}$, where we recall that $K_\mu = -\Delta - \mu$.

\vspace*{1ex} \noindent $\bullet$\ Given a trial one-particle density
matrix $\gamma$ and a small number $\delta > 2 \mu U^{-1}$, we
introduce the corresponding particle density $\rho(x) := \gamma_{x,x}$
and define the regions $\Omega := \{ x | \rho(x) < \delta \}$ and
$\Omega^c := \{ x | \rho(x) \geq \delta \}$ of low and high density onto
which we project by $P_\Om = \sum_{x \in \Om} | x \ra\la x |$ and
$P_\Om^\perp = \one - P_\Om$, respectively.

\vspace*{1ex} \noindent $\bullet$\ We then use the fact that $\gamma$
is mostly localized in the high density region $\Om^c$. This leads us
to estimate the kinetic energy $\Tr\{ - \Delta P_\Om \gamma P_\Om \}$
in $\Om$ by zero and $\Tr\{ - \Delta P_\Om^\perp \gamma P_\Om^\perp
\}$ in $\Om^c$ by the kinetic energy of the free Fermi gas in $\Om^c$.
The localization error is of order of a small constant times the
volume $|\pOm|$ of the boundary of $\Om$. In
\ul{Lemma~\ref{lem-IIa.1}} we give the exact formulation of the bound
which we use to estimate the term $\Tr\{ K_\mu \gamma \}$ in
$\tcE_{\mu,U}^\hfz(\gamma)$.

\vspace*{1ex} \noindent $\bullet$\ For the analysis of the term $\Tr\{
[K_\mu + U \rho]_- \}$ in $\tcE_{\mu,U}^\hfz(\gamma)$, we use the fact
that $\Om^c$ is a classically forbidden region, because $-\mu + U \rho
\geq -\mu + U \delta \geq \mu$ in $\Om^c$. So, as shown in
\ul{Lemma~\ref{lem-IIb.1}}, we can replace $\Tr\{ [K_\mu + U \rho]_-
\}$ by $\Tr\{ [P_\Om ( K_\mu + U \rho ) P_\Om]_- \}$, up
to localization errors of order of a small constant times $|\pOm|$.

\vspace*{1ex} \noindent $\bullet$\ We then pick a (large, but fixed)
number $M > 1$ and further split up the low density region $\Om$ into
the subset $\Om_1$ of those points in $\Om$ that are at most at
distance $2M$ away from the boundary $\pOm$ and the {\it bulk}
$\Om_2 \subset \Om$ of points of distance $2M$ or more to $\pOm$.
The contribution of $\Om_1$ turns out to be negligible
because $\Om_1$ contains at most $(4M+1)^d |\pOm|$ points, and the
density is low in $\Om_1 \subseteq \Om$.

\vspace*{1ex} \noindent $\bullet$\ The estimate of the region $\Om_2
\ni x$ then uses the lower bound on the spatial density $\one[K_\mu + U
\rho < 0](x,x)$ of the projection onto the negative eigenvalues of
$K_\mu + U \rho$ (actually, $\trho$ instead of $\rho$), which we
derive in \ul{Lemma~\ref{lem-IIc.1}}

\vspace*{1ex} \noindent $\bullet$\ Adding up the estimates derived so
far, we finally observe that $\tcE_{\mu,U}^\hfz(\gamma)$ is bounded
below by $\Tr\{ [P_\Om K_\mu P_\Om]_-\} + \Tr\{ [P_\Om^\perp K_\mu
P_\Om^\perp ]_-\} - \eta |\pOm| =: Y - \eta |\pOm|$, where $\eta >0$
becomes small when $U \gg 1$ and $\delta>0$ is properly chosen. In
\ul{Lemma~\ref{lem-IId.1}}, we reproduce the result from
\cite{FreericksLiebUeltschi2002,Goldbaum2003} that $Y$ can be
estimated from below by $\Tr\{ [ K_\mu ]_- \} + \alpha |\pOm|$, where
$\alpha >0$ depends only on $\mu$. In other words, the introduction of
a domain wall at $\pOm$ drives up the energy by $\alpha |\pOm|$, which
dominates $\eta |\pOm|$, provided $\eta$ is small. This establishes
that $\tcE_{\mu,U}^\hfz(\gamma) \geq \Tr\{ [ K_\mu ]_- \} + (\alpha -
\eta) |\pOm|$, which implies the claim.

\vspace*{1ex}

To carry out the proof in detail, we start with the observation that
the minimization over {\it two} one-particle density matrices in
(\ref{eq-I.7}) can actually be reduced to the minimization over only
{\it one} one-particle density matrix. To see this, we observe that
\begin{equation} \label{eq-II.1}
\sum_{x \in \Lambda} \rho_\ua(x) \, \rho_\da(x) 
\ = \ 
\Tr\{ \rho_\ua \, \gamma_\da \} \comma
\end{equation}
where $\rho_\ua$ acts as a multiplication operator, 
$\left(\rho_\ua f \right)(x) := \rho_\ua(x) f(x)$. Thus we have
\begin{eqnarray} 
E_{\mu,U}^\hfz 
& = & 
\min_{0 \leq \gamma_\ua \leq 1}\Big[
\Tr\{ K_\mu \, \gamma_\ua \} \; + \;
\min_{0 \leq \gamma_\da \leq 1}\big(
\Tr\{ (K_\mu + U \rho_\ua) \, \gamma_\da \} \big) \Big]   
\label{eq-II.2a} \\
& = & 
\min_{0 \leq \gamma_\ua \leq 1}\Big(
\Tr\{ K_\mu \, \gamma_\ua \} \; + \;
\Tr\{ [K_\mu + U \rho_\ua]_- \} \Big) \  . \label{eq-II.2b}
\end{eqnarray}
(Recall that $K_\mu = -\Delta - \mu$.) 
In other words, we have done the minimization over $ \gamma_\da$ in
(\ref{eq-II.2a}) by taking $ \gamma_\da$ to be the projection onto the
negative eigenspaces of $ K_\mu + U \rho_\ua$. Thus, as our
minimization principle over only one $\gamma$, we obtain the
following.
\begin{eqnarray} \label{eq-II.3a}
E_{\mu,U}^\hfz 
& = &
\min\big\{ \tcE_{\mu,U}^\hfz (\gamma) \; \big|
\ 0 \leq \gamma \leq 1 \big\} \comma
\\ \label{eq-II.3b}  
\tcE_{\mu,U}^\hfz(\gamma) 
& := &
\Tr\{ K_\mu \, \gamma \} \; + \;
\Tr\{ [K_\mu + U \rho]_- \} \comma
\end{eqnarray}
where $\rho(x) := \gamma_{x,x}$. From now on $\gamma$, with $0 \leq
\gamma \leq 1$, is an arbitrary, but fixed one-particle density
matrix, for which we bound  $\tcE_{\mu,U}^\hfz(\gamma)$ from below.
(An upper bound that agrees with Theorem~\ref{thm-I.1} is readily
obtained simply by choosing the variational function consisting of the
unperturbed Fermi sea with all particles spin-up or all spin-down.)

For the next step of the proof we introduce a small number 
$\delta > 2 \mu U^{-1}$, whose precise value will be chosen in the 
final step of the proof. Given a one-particle density matrix 
$0 \leq \gamma \leq 1$ with corresponding density 
$\rho(x) := \gamma_{x,x}$, we write the lattice 
$\Lambda = \Om \cup \Om^c$ as a union of two disjoing subsets
of $\Lambda$ in the following way.
\begin{eqnarray} \label{eq-II.4}
\Om \
& := & 
\big\{ x \in \Lambda \; \big| \ \rho(x) < \delta \big\} \comma
\\ \label{eq-II.5} 
\Om^c
& := & 
\big\{ x \in \Lambda \; \big| \ \rho(x) \geq \delta \big\} \ .
\end{eqnarray}
These are the regions of low and high density,
respectively. We define the boundary $\pOm$ of $\Om$ by
\begin{equation} \label{eq-II.6}
\pOm \ := \ 
\big\{ x \in \Om \; \big| \ \dist_1(x, \Om^c) = 1 \big\} \comma
\end{equation}
where $\dist_1(x, A) $ is the length of (number of bonds in) a
shortest path joining $x$ and some point in $y \in A$.  Another useful
notion of distance which we shall use is $\dist_\infty (x, A) $, which
is defined by the condition that $2 \, \dist_\infty(x, A) + 1$ is the
sidelength of the smallest cube centered at $x$ that intersects $A$.
When $A$ is a single point $y$ these distances are denoted by
$|x-y|_1$ and $|x-y|_\infty$.

We define  $P_\Om$, $P_{\Om^c} = P_\Om^\perp$, and $P_{\pOm}$ to be
the orthogonal projections onto $\Om$, $\Om^c$, and $\pOm$, 
respectively, where the projection onto an arbitrary set 
$A \subseteq \Lambda$ is given by 
\begin{equation} \label{eq-II.7}
(P_A f)(x) =\ \left\{
\begin{array}{cc}
f(x)  & \mbox{for $x \in A$,} \\
0  &  \mbox{for $x \notin A $.} \\    
\end{array}     \right.
\end{equation}
We further set 
\begin{equation} \label{eq-II.8}
\trho(x) \ := \ \left\{
\begin{array}{cc}
\rho(x) \comma & \mbox{for $x \in \Om^c$,} \\
\min\big\{\frac{\mu}{2 \, U} \, , \; \rho(x) \big\} \comma & 
\mbox{for $x \in \Om$,} \\
\end{array} \right. 
\end{equation}
and observe that $\trho(x) \leq \rho(x)$, for all $x \in \Lambda$,
which implies that
\begin{equation} \label{eq-II.9}
\tcE_{\mu,U}^\hfz (\gamma) 
\ \geq \
\Tr\{ K_\mu \, \gamma \} \; + \;
\Tr\{ [K_\mu + U \trho]_- \} \period
\end{equation}
For brevity,  we define  
$M := M_*(\mu) := 12 \, (\frac{4d}{\mu})^2 $ 
and  note that, by
assumption, $L $ obeys $L \geq 2 M$. We further decompose
$\Om$ into two disjoint subsets $\Om_1$ and $\Om_2$ defined by
\begin{eqnarray} \label{eq-II.10}
\Om_1
& := & 
\big\{ x \in \Om \; \big| \ \dist_\infty(x, \Om^c) \leq 2M \big\} \comma
\\ \label{eq-II.11} 
\Om_2
& := & 
\big\{ x \in \Om \; \big| \ \dist_\infty(x, \Om^c) > 2M \big\} \period
\end{eqnarray}
We observe that the $\ell^\infty$-distance of the points in $\Om_1$
to the boundary $\pOm$ of $\Om$ is less or equal to $2M$, so
$\Om_1 \subseteq \pOm + Q(2M)$, where 
$Q(\ell) = \{-\ell, \ldots, \ell\}^d + L \ZZ^d$. Hence
\begin{equation} \label{eq-II.12}
|\Om_1| \ \leq \ |\pOm| \cdot |Q(2M)| 
\ = \ 
(4M+1)^d \cdot |\pOm| \comma
\end{equation}
and therefore
\begin{equation} \label{eq-II.13}
\sum_{x \in \Om} \rho(x) 
\ = \ 
\sum_{x \in \Om_1} \rho(x) \: + \: 
\sum_{x \in \Om_2} \rho(x) 
\ \leq \ 
(4M+1)^d \, \delta \, |\pOm| \: + \: 
\sum_{x \in \Om_2} \rho(x) \comma
\end{equation}
since $\rho \leq \delta$ on $\Om$. 
Eq.~(\ref{eq-II.13}) and Lemma~\ref{lem-IIa.1} yield
\begin{eqnarray} \label{eq-II.14}
\Tr\{ K_\mu \, \gamma \}
& \geq &
\Tr\big\{ [ P_\Om^\perp K_\mu P_\Om^\perp ]_- \big\}
\\ \nonumber 
& &
\; - \; \big( 4d \, \delta^{1/2} + \mu \, (4M+1)^d \, \delta \big) \, |\pOm|  
\; - \; \mu \sum_{x \in \Om_2} \rho(x) \period
\end{eqnarray}
Next, we apply Lemma~\ref{lem-IIb.1} which asserts 
\begin{equation} \label{eq-II.15}
\Tr\{ [K_\mu + U \trho]_- \}
\ \geq \
\Tr\big\{ [ P_\Om (K_\mu + U \trho) P_\Om ]_- \big\}
\; - \; \frac{8d^2}{U \, \delta} \, |\pOm| \period
\end{equation}
Denoting by $\chi := \one[ P_\Om (K_\mu + U \trho) P_\Om < 0]$ the
orthogonal projection onto the subspace of negative eigenvalues of
$P_\Om (K_\mu + U \trho) P_\Om$ and $\rho_\chi(x) := \chi_{x,x}$ its
diagonal matrix element, we observe that
\begin{eqnarray} \label{eq-II.16}
\Tr\big\{ [ P_\Om (K_\mu + U \trho) P_\Om ]_- \big\}
& = &
\Tr\big\{ P_\Om (K_\mu + U \trho) P_\Om \, \chi \big\}
\\ \nonumber 
& = &
\Tr\big\{ P_\Om \, K_\mu \, P_\Om \, \chi \big\}
\; + \;
U \sum_{x \in \Om} \rho_\chi(x) \, \trho(x) \period
\end{eqnarray}
By Lemma~\ref{lem-IIc.1}, the density $\rho_\chi$ is bounded
below on $\Om_2$ by the universal constant $\kappa(\mu) >0$
defined in (\ref{eq-IIc.3}). Therefore
\begin{equation} \label{eq-II.17}
\Tr\{ [K_\mu + U \trho]_- \}
\ \geq \
\Tr\big\{ [P_\Om \, K_\mu \, P_\Om]_- \big\}
\; - \; 
\frac{8d^2}{U \, \delta} \, |\pOm| 
\; + \;
\kappa(\mu) \, \sum_{x \in \Om_2} U \, \trho(x) \period
\end{equation}
Adding up (\ref{eq-II.14}) and (\ref{eq-II.17}), we obtain 
\begin{eqnarray} \label{eq-II.18}
\tcE_{\mu,U}^\hfz (\gamma) 
& \geq &
\Tr\big\{ [P_\Om \, K_\mu \, P_\Om]_- \big\}
\; + \;
\Tr\big\{ [ P_\Om^\perp K_\mu P_\Om^\perp ]_- \big\}
\nonumber \\
& & 
\; - \; 
\Big\{ 4d \, \delta^{1/2} + \mu \, (4M+1)^d \, \delta + 
\frac{8d^2}{U \, \delta} \Big\} \, |\pOm|  
\nonumber \\
& &
\; + \; 
\sum_{x \in \Om_2} 
\big\{ \kappa(\mu) \, U \, \trho(x)- \mu \, \rho(x) \big\} \comma
\end{eqnarray}
and Lemma~\ref{lem-IId.1} further yields
\begin{eqnarray} \label{eq-II.18,5}
\tcE_{\mu,U}^\hfz (\gamma) 
\; - \;
\Tr\big\{ [K_\mu]_- \big\}
& \geq &
\Big\{ \alpha(\mu) - 4d \, \delta^{1/2} - \mu \, (4M+1)^d \, \delta - 
\frac{8d^2}{U \, \delta} \Big\} \, |\pOm|  
\nonumber \\
& &
\; + \; 
\sum_{x \in \Om_2} 
\big\{ \kappa(\mu) \, U \, \trho(x)- \mu \, \rho(x) \big\} \period
\end{eqnarray}
We choose
\begin{equation} \label{eq-II.19}
\delta 
\ := \ 
\delta_*(\mu) 
\ = \
\min\Big\{ \frac{\alpha(\mu)^2}{(12 d)^2} \comma \ \ 
\frac{\alpha(\mu)}{3\mu (4M+1)^d} \comma \ \ \frac{\kappa(\mu)}{2} \Big\}
\comma
\end{equation}
and we observe that if 
\begin{equation} \label{eq-II.20}
U \ \geq \ U_*\big( \mu, \alpha(\mu) \big) 
\ = \ 
\max\Big\{ \frac{2\mu}{\delta_*(\mu,\alpha(\mu))} \comma \ \ 
\frac{24 \, d^2}{\alpha(\mu) \, \delta_*(\mu,\alpha(\mu))} \Big\} 
\end{equation}
then our choice for $\delta$ fulfills the requirement 
$\delta > 2 \mu U^{-1}$. Moreover, Eqs.~(\ref{eq-II.19}) and 
(\ref{eq-II.20}) imply that
\begin{equation} \label{eq-II.21}
4d \, \delta^{1/2} + \mu \, (4M+1)^d \, \delta + \frac{8d^2}{U \, \delta} 
\ \leq \ 
\frac{\alpha(\mu)}{3} + \frac{\alpha(\mu)}{3} + \frac{\alpha(\mu)}{3}  
\ \leq \ 
\alpha(\mu) \period
\end{equation}
We further set $\Om_2' := \{x \in \Om_2| \; \rho(x) \leq \frac{\mu}{2U} \}$
and $\Om_2'' := \{x \in \Om_2| \; \frac{\mu}{2U} < \rho(x) \leq \delta \}$,
so $\Om_2$ is the disjoint union of $\Om_2'$ and $\Om_2''$, and
by the definition (\ref{eq-II.8}) of $\trho$, we have that
\begin{eqnarray} \label{eq-II.22}
\lefteqn{
\sum_{x \in \Om_2} 
\big\{ \kappa(\mu) \, U \, \trho(x) - \mu \, \rho(x) \big\}
}
\\ \nonumber 
& \geq & 
\sum_{x \in \Om_2'} 
\{ \kappa(\mu) \, U - \mu \} \, \rho(x)
\; + \;
\sum_{x \in \Om_2''} \frac{\mu}{2} \: 
\big\{ \kappa(\mu) - 2 \delta \big\}
\ \; \geq \ \; 0 \comma
\end{eqnarray}
since $\delta \leq \frac{1}{2} \kappa(\mu)$ and $U \geq
2\mu/\delta_*(\mu,\alpha(\mu)) \geq \mu / \kappa(\mu)$.
Eqs.~(\ref{eq-II.21}) and (\ref{eq-II.22}) insure that the right side
of (\ref{eq-II.18,5}) is nonnegative, which immediately implies
Theorem~\ref{thm-I.1}.

Theorem~\ref{thm-I.1a} is obtained by substituting the explicit value
of $\alpha(\mu)$ from (\ref{eq-IId.4}) into (\ref{eq-II.20}) and using
$L_* (\mu)$ from (\ref{eq-IId.4}) .  \hfill {\bf QED}

\secct{Auxiliary Lemmas}
\label{sec-III}


In this section we state and prove the lemmas used in the proof of Theorems \ref{thm-I.1} and  \ref{thm-I.1a} 
in Section \ref{sec-II}.

\subsection{The Region $\Om^c$ of High Density} 
\label{subsec-IIa}
In this subsection, we estimate $\Tr\{ K_\mu \, \gamma \}$
from below. We are guided by the intuition that $\gamma$ is
essentially localized on $\Om^c$.
\begin{lemma} \label{lem-IIa.1}
\begin{equation} \label{eq-IIa.1}
\Tr\{ K_\mu \, \gamma \}
\ \geq \
\Tr\big\{ [ P_\Om^\perp K_\mu P_\Om^\perp ]_- \big\}
\; - \; 4d \, \delta^{1/2} \, |\pOm|  
\; - \; \mu \sum_{x \in \Om} \rho(x) \period
\end{equation}
\end{lemma}
\Proof
Inserting $\one = P_\Om + P_\Om^\perp$ into 
$\Tr\{ K_\mu \, \gamma \}$, we obtain
\begin{eqnarray} \label{eq-IIa.2}
\Tr\{ K_\mu \, \gamma \}
& = & 
\Tr\{ K_\mu \, P_\Om^\perp \, \gamma \, P_\Om^\perp \} 
\; - \; 
2 \re \, \Tr\{ P_\Om^\perp \, \Delta \, P_\Om \, \gamma \} 
\; + \; 
\Tr\{ K_\mu \, P_\Om \, \gamma \, P_\Om \} 
\nonumber \\[1ex] 
& \geq & 
\Tr\big\{ [ P_\Om^\perp K_\mu P_\Om^\perp ]_- \big\}
\; - \; 
2 \sum_{x \in \Om, y \in \Om^c} \Delta_{x,y} \, |\gamma_{y,x}|
\; - \; 
\mu \, \Tr\{ P_\Om \, \gamma \, P_\Om \} 
\nonumber \\ 
& = & 
\Tr\big\{ [ P_\Om^\perp K_\mu P_\Om^\perp ]_- \big\}
\; - \; 
2 \sum_{x \in \pOm, y \in \Om^c} \Delta_{x,y} \, |\gamma_{y,x}|
\; - \; 
\mu \, \sum_{x \in \Om} \rho(x) \comma \hspace{5mm}
\nonumber \\ 
& & 
\end{eqnarray}
where we use that $-\Delta \geq 0$, that 
$P_\Om^\perp \Delta P_\Om = P_\Om^\perp \Delta P_\pOm$, and that 
$0 \leq \gamma \leq 1$. The latter also implies that 
$\rho(y) = \gamma_{y,y} \leq 1$, for all $y \in \Lambda$. 
Thus, if $x \in \pOm$ and $y \in \Om^c$, the Cauchy-Schwarz inequality 
yields $|\gamma_{y,x}| \leq \sqrt{ \gamma_{y,y} \cdot \gamma_{x,x} \,}
\leq \delta^{1/2}$. Moreover, if $x \in \pOm$, $y \in \Om^c$, and
$\Delta_{x,y} \neq 0$, then $y$ is a neighbor of $x$, and 
we obtain
\begin{equation} \label{eq-IIa.3}
\sum_{x \in \pOm, y \in \Om^c} \Delta_{x,y} \, |\gamma_{y,x}|
\ \leq \
\delta^{1/2} \, \sum_{x \in \pOm} \, \sum_{y \in \Lambda: \, |x-y|=1}
\ = \
2d \, \delta^{1/2} \, |\pOm| \comma
\end{equation}
which completes the proof of (\ref{eq-IIa.1}).
\hfill {\bf QED}

\subsection{Decoupling the High and Low Density Regions} 
\label{subsec-IIb}
%
This subsection is devoted to showing that
$\Tr\{ [K_\mu + U \trho]_- \}$ essentially agrees with
the corresponding eigenvalue sum 
$\Tr\{ [ P_\Om(K_\mu + U \trho)P_\Om]_- \}$ for the 
operator localized on $\Om$, the reason being that $\Om^c$ is a 
classically forbidden region since 
$-\mu + U \trho \geq \frac{1}{2} U \delta > 0$ on $\Om^c$.
\begin{lemma} \label{lem-IIb.1}
\begin{equation} \label{eq-IIb.1}
\Tr\{ [K_\mu + U \trho]_- \}
\ \geq \
\Tr\big\{ [ P_\Om (K_\mu + U \trho) P_\Om ]_- \big\}
\; - \; \frac{8d^2}{U \, \delta} \, |\pOm| \period
\end{equation}
\end{lemma}
\Proof
We wish to apply of the Feshbach projection method. To this end, we
first observe the following quadratic form bound,
\begin{equation} \label{eq-IIb.2}
P_\Om^\perp (K_\tmu + U \trho) P_\Om^\perp
\ \geq \
P_\Om^\perp (U \trho - \tmu ) P_\Om^\perp
\ \geq \
\frac{1}{2} U \, \delta \, P_\Om^\perp \comma
\end{equation}
for any $\tmu \in [0,\mu]$, since $\trho \geq \delta$ on $\Om^c$ 
and $\delta \geq 2 \mu U^{-1}$.
Thus, $P_\Om^\perp (K_\tmu + U \trho) P_\Om^\perp$ is positive
and invertible on $\Ran P_\Om^\perp$, and moreover, we have that
\begin{equation} \label{eq-IIb.3a}
P_\Om \, \Delta \, P_\Om^\perp 
\big[ P_\Om^\perp (K_\tmu + U \trho) P_\Om^\perp \big]^{-1} 
P_\Om^\perp \, \Delta \, P_\Om 
\ \leq \
\frac{2}{U \, \delta} \, 
P_\pOm \, \Delta \, P_\Om^\perp \, \Delta \, P_\pOm \period
\end{equation}
For $y \in \Om^c$ and $f \in \CC^\Lambda$, the Cauchy-Schwarz 
inequality implies that
\begin{eqnarray} \label{eq-IIb.4,1}
\lefteqn{
\la f | \; P_\pOm \, \Delta \, \one_y \, \Delta \, P_\pOm \, f \ra
\ \; = \; \
| (\Delta P_\pOm f)[y] |^2 
\ \; = \; \
\Big| \sum_{x \in \pOm, |x-y|_1 = 1} f(x) \Big|^2
}
\\ \nonumber 
& \leq & 
\Big( \sum_{x \in \pOm, |x-y|_1 = 1} \!\! |f(x)|^2 \Big)
\Big( \sum_{x \in \Lambda, |x-y|_1 = 1} \!\! 1 \Big)
\ \; = \; \
2d \: \sum_{x \in \pOm, |x-y|_1 = 1} \!\! |f(x)|^2 \comma
\end{eqnarray}
which, by summing over all $y \in \Om^c$, yields
\begin{eqnarray} \label{eq-IIb.4,2}
\la f | \; P_\pOm \, \Delta \, P_\Om^\perp \, \Delta \, P_\pOm \, f \ra
& = &
\sum_{y \in \Om^c}
\la f | \; P_\pOm \, \Delta \, \one_y \, \Delta \, P_\pOm \, f \ra
\\ \nonumber 
& \leq & 
2d \: \sum_{x \in \pOm} \Big\{ |f(x)|^2 \cdot
\Big( \sum_{y \in \Lambda, |x-y|_1 = 1} \!\! 1 \Big) \Big\}
\\ \nonumber 
& \leq & 
4d^2 \: \sum_{x \in \pOm} |f(x)|^2 
\ \; = \; \
4d^2 \: \la f | \; P_\pOm \, f \ra \period
\end{eqnarray}
(We thank D.~Ueltschi for pointing out (\ref{eq-IIb.4,1})--(\ref{eq-IIb.4,2}) 
to us.) We conclude that
\begin{equation} \label{eq-IIb.3b}
P_\Om \, \Delta \, P_\Om^\perp 
\big[ P_\Om^\perp (K_\tmu + U \trho) P_\Om^\perp \big]^{-1} 
P_\Om^\perp \, \Delta \, P_\Om 
\ \leq \
\frac{8d^2}{U \, \delta} \, P_\pOm \period
\end{equation}

The invertibility of $P_\Om^\perp (K_\tmu + U \trho + e) P_\Om^\perp$ on 
$\Ran P_\Om^\perp$ implies the applicability of the Feshbach map,
for any $e \in [0,\mu]$. I.e., for any $e \in [0,\mu]$, 
\begin{eqnarray} \label{eq-IIb.5}
F(e) & := &
F_{P_\Om}[K_\mu + e + U \trho] - e \, P_\Om 
\\ \nonumber 
& = & 
P_\Om (K_\mu + U \trho) P_\Om 
\; - \;
P_\Om \, \Delta \, P_\Om^\perp 
\big[ P_\Om^\perp (K_\mu + e + U \trho) P_\Om^\perp \big]^{-1} 
P_\Om^\perp \, \Delta \, P_\Om 
\end{eqnarray}
is a well-defined matrix on $\Ran P_\Om$, 
and the isospectrality of the Feshbach map guarantees that
$-e \in [-\mu,0)$ is a negative eigenvalue of $K_\mu + U \trho$
of multiplicity $m(e)$ if and only if $-e$ is an
(nonlinear) eigenvalue of $F(e)$, i.e., if the kernel
of $F(e)+e$, as a subspace of $\Ran P_\Om$, has dimension $m(e)$.
Note that $F$ is monotonically increasing, as a quadratic form, 
in $e >0$. In particular,
\begin{equation} \label{eq-IIb.6}
F(e) \ \geq \ F(0) 
\ \geq \ 
P_\Om (K_\mu + U \trho) P_\Om \; - \; 
\frac{8d^2}{U \, \delta} \, P_\pOm \comma
\end{equation}
additionally taking (\ref{eq-IIb.3b}) into account.

We claim that, for all $\lambda \in (0,\infty)$, the number of 
eigenvalues of $K_\mu + U \trho$ below $-\lambda$ is smaller than
the number of negative eigenvalues of $F(\lambda) + \lambda$, 
\begin{equation} \label{eq-IIb.7}
\Tr\big\{ \one[K_\mu + U \trho < -\lambda] \big\}
\ \leq \
\Tr_\Om\big\{ \one[F(\lambda) + \lambda < 0] \big\} \comma
\end{equation}
where $\Tr_\Om$ denotes the trace on $\Ran P_\Om$.
Both sides of Eq.~(\ref{eq-IIb.7}) are zero and thus fulfill the
claimed inequality, for $\lambda \geq \mu$. Assume that
(\ref{eq-IIb.7}) is violated, for some $\lambda \in (0,\infty)$, i.e.,
that $\lambda_* := \inf\{\lambda \in (0,\infty) \, | \;
\mbox{Eq.~(\ref{eq-IIb.7}) holds true}\} >0$. We show that this
assumption leads to a contradiction. Obviously, $-\lambda_*$ must be
an eigenvalue of $K_\mu + U \trho$, and hence also of $F(\lambda_*)$,
of multiplicity $m(\lambda_*) \geq 1$, because only then the left or
the right side of (\ref{eq-IIb.7}) changes (increases, in fact).
Moreover, Eq.~(\ref{eq-IIb.7}) holds true for $\lambda = \lambda_*$
itself, i.e., the infimum in the definition of $\lambda_*$ is a
minimum. Hence, for all sufficiently small $\eps >0$, the definition
of $\lambda_*$ and the monotony of $F(e)$ in $e$ yield
\begin{eqnarray} \label{eq-IIb.8}
\Tr\big\{ \one[K_\mu + U \trho < -\lambda_*] \big\}
& \leq &
\Tr_\Om\big\{ \one[F(\lambda_*) + \lambda_* < 0] \big\} 
\\ \label{eq-IIb.9}
\Tr\big\{ \one[K_\mu + U \trho < -\lambda_* + \eps] \big\}
& > &
\Tr_\Om\big\{ \one[F(\lambda_* -\eps) + \lambda_* - \eps < 0] \big\} 
\nonumber \\
& \geq &
\Tr_\Om\big\{ \one[F(\lambda_*) + \lambda_* - \eps < 0] \big\} \period
\end{eqnarray}
Choosing $\eps >0$ so small that $-\lambda_*$ is the only eigenvalue
of $K_\mu + U \trho$ in the interval $[-\lambda_*, -\lambda_* + \eps]$, 
we hence obtain
\begin{eqnarray} \label{eq-IIb.10}
m(\lambda_*) & = &
\Tr\big\{ \one[0 \leq K_\mu + U \trho + \lambda_* < \eps] \big\}
\nonumber \\
& = & 
\Tr\big\{ \one[K_\mu + U \trho < -\lambda_* + \eps] \big\}
\; - \;
\Tr\big\{ \one[K_\mu + U \trho < -\lambda_*] \big\} 
\nonumber \\
& > &
\Tr_\Om\big\{ \one[F(\lambda_*) + \lambda_* < \eps] \big\} 
\; - \;
\Tr_\Om\big\{ \one[F(\lambda_*) + \lambda_* < 0] \big\} 
\nonumber \\
& = & 
\Tr_\Om\big\{ \one[0 \leq F(\lambda_*) + \lambda_* < \eps] \big\} 
\ \; = \ \;
m(\lambda_*) \comma
\end{eqnarray}
arriving at a contradiction, which proves (\ref{eq-IIb.7}), for all
$\lambda \in (0,\infty)$. From (\ref{eq-IIb.7}) and (\ref{eq-IIb.6}),
we finally conclude 
\begin{eqnarray} \label{eq-IIb.11}
\Tr\{ [K_\mu + U \trho]_- \}
& = &
- \int_0^\infty 
\Tr\big\{ \one[K_\mu + U \trho < -\lambda] \big\} \: d\lambda
\nonumber \\
& \geq &
- \int_0^\infty 
\Tr_\Om\big\{ \one[F(\lambda) + \lambda < 0] \big\} \period
\nonumber \\
& \geq &
- \int_0^\infty 
\Tr_\Om\big\{ \one[F(0) + \lambda < 0] \big\} 
\\ \nonumber 
& = &
\Tr\{ [F(0)]_- \}
\ \; = \ \;
\Tr_\Om\{ [F(0)]_- \}
\\ \nonumber 
& \geq &
\Tr\big\{ [ P_\Om (K_\mu + U \trho) P_\Om ]_- \big\}
\; - \; 
\frac{8d^2}{U \, \delta} \, \Tr\{P_\pOm\} \period
\end{eqnarray}
which is the assertion of Lemma~\ref{lem-IIb.1}.
\hfill {\bf  QED}

\subsection{The Electron Density in the Bulk} 
\label{subsec-IIc}
%
In this subsection we consider the spectral projection 
\begin{equation} \label{eq-IIc.1}
\chi \ := \ 
\one\big[P_\Om \, (K_\mu + U \trho) \, P_\Om < 0 \big]
\ = \ 
\one\big[P_\Om \, (-\Delta - \mu + U \trho) \, P_\Om < 0 \big] 
\end{equation}
of $P_\Om \, (-\Delta - \mu + U \trho) \, P_\Om$ onto its negative 
eigenvalues. Writing $\Delta_\Om := P_\Om \, \Delta \, P_\Om$,
i.e., $(\Delta_\Om)_{x,y} = \Delta_{x,y}$, for $x,y \in \Om$, and
$= 0$, otherwise, and 
$V \equiv \sum_{x \in \Om} V(x) \cdot \one_x := \mu P_\Om - U \trho P_\Om$, 
we have that
\begin{equation} \label{eq-IIc.2}
\chi \ = \ \one[ -\Delta_\Om - V < 0 ]
\hspace{4mm} \mbox{and} \hspace{4mm}
\forall x \in \Om: \quad 
\frac{1}{2} \mu \; \leq \; V(x) \; \leq \; \mu \comma
\end{equation}
due to the definition (\ref{eq-II.7}) of $\trho$. Naive semiclassical
intuition tells us that, for $x \in \Om$, the particle density
$\rho_\chi(x) := \chi_{x,x}$ corresponding to the one-particle density
matrix $\chi$ should be bounded below by the particle density of the
Fermi gas given by the one-particle density matrix 
$\one[-\Delta < \mu/2]$. The purpose of this
subsection is to prove such a bound (up to a constant factor) 
where it can be expected to hold, namely, for those points $x$
that are sufficiently far away from the boundary of $\Om$.
\begin{lemma} \label{lem-IIc.1}
Let $0< \mu \leq 4d$, define
$M := M_* := 
12 (\frac{4d}{\mu})^2 $.
Suppose that $L $ obeys $L \geq 2 M$ and that 
$x \in \Om$, with $\dist_\infty(x, \pOm) > 2M$. Then 
\begin{equation} \label{eq-IIc.3}
\rho_\chi (x) 
\ \geq \
\kappa(\mu)
\ := \
\frac{\mu^d}{4^{2d+1} \, e^d \, d^d} \;
\Big[1 + 2 \, \ln(2) \, \big( d^{-1} +1 \big) + 
     \ln\big(4d \mu^{-1} \big) \Big]^{-2d} \period
\end{equation}
\end{lemma}
\Proof
For any $\beta >0$, we note that the map $\RR^\Om \to \RR$,
$W \mapsto (e^{-\beta (-\Delta_\Om - W)})_{x,x}$ is monotonically
increasing in $W$. Namely, as $T_\Om = P_\Om T P_\Om$ has nonnegative
matrix elements, so does $e^{\eps \Delta_\Om}$,
\begin{equation} \label{eq-IIc.3,1}
\big( e^{\eps \Delta_\Om} \big)_{w,z}
\ = \
e^{-2d\eps} \, \big( e^{\eps T_\Om} \big)_{w,z}
\ = \
e^{-2d\eps} \, \sum_{k=0}^\infty \frac{\eps^k}{k!} \big( T_\Om^k \big)_{w,z}
\ \geq \ 0 \comma
\end{equation}
for all $w,z \in \Om$. So, if $n$ is an integer  and $W, \tW \in \RR^\Om$ with 
$W(z) \leq \tW(z)$, for all $z \in \Om$, then we have that
\begin{eqnarray} \label{eq-IIc.3,2}
\Big( \big[e^{\beta \Delta_\Om/n} \, e^{\beta W/n} \big]^n \Big)_{z_0, z_n}
& = &
\sum_{z_1, \ldots, z_{n-1} \in \Om}
\Big\{ \prod_{j=1}^n \big( e^{\beta \Delta_\Om/n} \big)_{z_{j-1},z_j} 
\: e^{\beta W(z_j)/n} \Big\}
\nonumber \\
& \leq &
\sum_{z_1, \ldots, z_{n-1} \in \Om}
\Big\{ \prod_{j=1}^n \big( e^{\beta \Delta_\Om/n} \big)_{z_{j-1},z_j} 
\: e^{\beta \tW(z_j)/n} \Big\}
\nonumber \\
& = &
\Big( \big[e^{\beta \Delta_\Om/n} \, e^{\beta \tW/n} \big]^n \Big)_{z_0, z_n}
\comma
\end{eqnarray}
for all $z_0, z_n \in \Om$. Setting $z_0 := z_n := x \in \Om$ and
taking the limit $n \to \infty$, the Lie-Trotter product formula and 
Eq.~(\ref{eq-IIc.3,2}) imply that
\begin{equation} \label{eq-IIc.3,3}
\big( e^{-\beta (-\Delta_\Om - W)} \big)_{x,x}
\ \leq \
\big( e^{-\beta (-\Delta_\Om - \tW)} \big)_{x,x} \comma
\end{equation}
indeed. In particular,
\begin{equation} \label{eq-IIc.3,4}
e^{\beta \mu/2} \, \big( e^{\beta \Delta_\Om} \big)_{x,x}
\ \leq \
\big( e^{-\beta (-\Delta_\Om - V)} \big)_{x,x} \comma
\end{equation}
since $V \geq \frac{1}{2} \mu$ on $\Om$. On the other hand, 
$-\Delta_\Om - V \geq -\mu$ and 
$\chi^\perp (-\Delta_\Om - V) \chi^\perp \geq 0$, as quadratic forms. 
The spectral theorem thus implies that
\begin{eqnarray} \label{eq-IIc.3,5}
\chi \, e^{-\beta (-\Delta_\Om - V)} \, \chi 
& \leq &
\chi \, e^{\beta \mu} \, \chi 
\ = \
e^{\beta \mu} \, \chi \comma
\\  \label{eq-IIc.3,6}
\chi^\perp \, e^{-\beta (-\Delta_\Om - V)} \, \chi^\perp 
& \leq &
\chi^\perp \ \leq \ P_\Om \period
\end{eqnarray}
Putting together (\ref{eq-IIc.3,4}), (\ref{eq-IIc.3,5}), and 
(\ref{eq-IIc.3,6}), using that $\chi$ and $-\Delta_\Om -V$ commute, 
we arrive at 
\begin{eqnarray} \label{eq-IIc.4}
e^{\beta \mu/2} \, \big( e^{\beta \Delta_\Om} \big)_{x,x}
& \leq & 
\big( e^{-\beta (-\Delta_\Om - V)} \big)_{x,x}
\nonumber \\
& = & 
\big( \chi \, e^{-\beta (-\Delta_\Om - V)} \, \chi \big)_{x,x} \; + \; 
\big( \chi^\perp \, e^{-\beta (-\Delta_\Om - V)} \, \chi^\perp \big)_{x,x} 
\nonumber \\
& \leq & 
e^{\beta \mu} \, \chi_{x,x} \; + \; 1 \period
\end{eqnarray}
Solving for $\rho_\chi(x) = \chi_{x,x}$, we therefore have
\begin{equation} \label{eq-IIc.5}
\rho_\chi (x) \ \geq \
e^{-\beta \mu/2} 
\big[ (e^{\beta \Delta_\Om})_{x,x} \: - \: e^{-\beta \mu/2} \big] \comma
\end{equation}
for any $x \in \Om$ and any $\beta >0$. 

Next, recall that $Q(M) = \{-M, \ldots, M\}^d + L \ZZ^d = 
\{ y \in \Lambda \, : \; |y|_\infty \leq M \}$
is the box of sidelength $2M+1$ centered at $0 \in \Lambda$. 
Since $\dist_\infty(x, \pOm) > 2M$, by assumption, we have that
\begin{equation} \label{eq-IIc.5,1}
Q(M) - z + x \ \subseteq \ \Om \comma
\end{equation}
for all $z \in Q(M)$. By Lemma~\ref{lem-IIc.2}, this inclusion
implies that
\begin{equation} \label{eq-IIc.5,2}
\big( \exp[\beta \Delta_\Om] \big)_{x,x} 
\ \geq \
\big( \exp[\beta \Delta_{Q(M) - z + x}] \big)_{x,x} 
\ = \
\big( \exp[\beta \Delta_{Q(M)}] \big)_{z,z} \comma 
\end{equation}
and by averaging this inequality over $z \in Q(M)$, we obtain
\begin{equation} \label{eq-IIc.5,3}
\big( \exp[\beta \Delta_\Om] \big)_{x,x} 
\ \geq \
\frac{1}{|Q(M)|} \sum_{z \in Q(M)} 
\big( \exp[\beta \Delta_{Q(M)}] \big)_{z,z} \period
\end{equation}
Now, we apply Lemma~\ref{lem-IIc.3} and arrive at
\begin{eqnarray} \label{eq-IIc.11}
\lefteqn{
\frac{1}{|Q(M)|} \sum_{z \in Q(M)} 
\big( \exp[\beta \Delta_{Q(M)}] \big)_{z,z} 
\ \geq \
\frac{e^{-d \beta / M}}{(2\pi)^d} 
\int_{[-\pi,\pi]^d} \exp[-\beta \, \om(k)] \: d^dk 
}
\nonumber \\
& \hspace{30mm} = &
\bigg( \frac{2 \, e^{-\beta / M}}{\pi} 
\int_{0}^{\pi/2} \exp[- 4 \, \beta \, \sin^2(t)] \: dt \bigg)^d
\comma \hspace{15mm}
\end{eqnarray}
where $\om(k) = \om(-k) = \sum_{\nu=1}^d 2 \big\{ 1 - \cos(k_\nu) \big\}
= \sum_{\nu=1}^d 4 \sin^2(k_\nu/2)$. Choosing $\beta \geq 1$, we observe that
$\frac{1}{\pi} \int_0^{\sqrt{\beta}\pi} e^{-t^2} dt \geq
\frac{1}{\pi} \int_0^{\pi} e^{-t^2} dt = 
\frac{1}{2 \sqrt{\pi}} \, \mathrm{erf}[\pi] \geq \frac{1}{4}$.
Using this and $\sin^2(t) \leq t^2$, we have the following estimate,
\begin{equation} \label{eq-IIc.12}
\frac{2 \, e^{-\beta / M}}{\pi} 
\int_{0}^{\pi/2} \exp[- 4 \, \beta \, \sin^2(t)] \: dt
\ \geq \ 
\frac{e^{-\beta / M}}{\beta^{1/2}} \cdot
\frac{1}{\pi} 
\int_{0}^{\sqrt{\beta}\pi} e^{-t^2} \, dt
\ \geq \
\frac{e^{-\beta / M}}{4 \, \beta^{1/2}} \period
\end{equation}
Inserting this estimate into (\ref{eq-IIc.11}) and then
the result in (\ref{eq-IIc.5,3}) and (\ref{eq-IIc.5}),
we obtain, for any $\beta \geq 1$, that
\begin{equation} \label{eq-IIc.13}
\rho_\chi (x) 
\ \geq \
e^{-\beta \mu/2} 
\Big[ \frac{e^{-d\beta / M}}{4^d \, \beta^{d/2}} 
\: - \: e^{-\beta \mu/2} \Big] 
\ = \
e^{-\tau d} \bigg[
\big(e^{1 - 2 d \tau /(M\mu)} \big)^{d/2} 
\Big(\frac{\mu}{16 e d} \cdot \frac{e^\tau}{\tau} \Big)^{d/2}
\: - \: 1 \bigg] \comma
\end{equation}
where $\tau := \beta \mu/d$. Note that if we require $\tau \geq 4$
then $\beta = \tau d /\mu \geq 1$, since $\mu \leq 4d$. We may thus
replace $\beta \in [1, \infty)$ by $\tau \in [4, \infty)$. Our goal is
to choose $\tau$ such that 
\begin{eqnarray} \label{eq-IIc.14}
\lefteqn{
\Big(\frac{\mu}{16ed} \cdot \frac{e^\tau}{\tau} \Big)^{d/2}
\ \geq \ 2
\hspace{3mm} \Longleftrightarrow \hspace{3mm} 
}
\\ \label
& & \label{eq-IIc.14,1}
\tau - \ln(\tau) 
\ \geq \ 
Y \; := \; 1 + 2\, \ln(2) \Big(\frac{1}{d} + 1 \Big) + 
\ln\Big(\frac{4d}{\mu}\Big) \period
\end{eqnarray}
Note that, due to $\mu \leq 4d$,
\begin{equation} \label{eq-IIc.14,2}
2.38 \ \leq \ 1 + 2 \, \ln(2) \ \leq \ Y 
\ \leq \ 
3 \, \ln\big(16 \, d \, \mu^{-1} \big) \period
\end{equation}
We choose $\tau := Y + 2 \ln(Y)$ and observe that 
$Y \geq 2.38$ insures $\tau \geq 4.11 \geq 4$, as required.
Moreover, with this choice, we have
\begin{eqnarray} \label{eq-IIc.15}
\tau - \ln(\tau) - Y
& = &
2 \ln(Y) \: - \: \ln\big[Y + 2 \ln(Y)\big]
\nonumber \\ 
& \geq &
\ln(Y) \: - \: \ln\big[1 + 2 \ln(Y) \, Y^{-1} \big]
\\ \nonumber 
& \geq &
\ln(Y) \: - \: 2 \ln(Y) \, Y^{-1} 
\ = \
2 \ln(Y) \Big( \frac{1}{2} \: - \: \frac{1}{Y} \Big) 
\ > \ 0 \comma
\end{eqnarray}
using that $\ln(1+\eps) \leq \eps$, for $\eps \geq 0$, and 
$Y \geq 2.38 > 2$. Thus, (\ref{eq-IIc.14,1}) and (\ref{eq-IIc.14}) 
hold true. Additionally, we observe that $Y \leq 3 \ln(\frac{16d}{\mu})$ and
\begin{equation} \label{eq-IIc.15,01}
\tau 
\ \leq \ 
Y \cdot \max_{r>0}\Big\{1 + 2\Big( \frac{\ln r}{r} \Big) \Big\}
\ = \ 
(1+ 2/e) Y 
\ \leq \ 2Y
\end{equation}
insures that $\frac{2d\tau}{\mu} \leq \frac{12 d}{\mu} \ln(\frac{16d}{\mu}) 
\leq 12 (\frac{4d}{\mu})^2 \leq M_* \leq M$. This, in turn, yields
\begin{equation} \label{eq-IIc.15,1}
\exp\Big[ 1 - \frac{2d\tau}{M \mu} \Big] 
\ \geq \ 1 \comma
\end{equation}
and by inserting (\ref{eq-IIc.15,1}) and (\ref{eq-IIc.14}) into 
(\ref{eq-IIc.13}), we arrive at
\begin{equation} \label{eq-IIc.16}
\rho_\chi (x) 
\ \geq \
e^{-\tau d} 
\ = \
\frac{\mu^d}{4^{2d+1} \, e^d \, d^d} \;
\Big[1 + 2 \, \ln(2) \, \big( d^{-1} +1 \big) + 
     \ln\big(4d \mu^{-1} \big) \Big]^{-2d} \period
\end{equation}
\hfill {\bf  QED }

\begin{lemma} \label{lem-IIc.2}
Let $A,B \subseteq \Lambda$, with $A \subseteq B$, and denote
$\Delta_A := P_A \Delta P_A$ and $\Delta_B := P_B \Delta P_B$.
For all $x \in A$ and all $\beta >0$, 
\begin{equation} \label{eq-IIc.30,1}
\big( \exp[\beta \, \Delta_A] \big)_{x,x} 
\ \leq \
\big( \exp[\beta \, \Delta_B] \big)_{x,x} \period
\end{equation}
\end{lemma}
\Proof
We first define the nearest-neighbor hopping matrix $T$ on
$\Lambda$ by $T_{w,z} := 1$ if $|w-z|_1 =1$ and $T_{w,z} := 0$, otherwise. 
For a given subset $C \subset \Lambda$, the matrix $T_C := P_C T P_C$ 
denotes the hopping matrix restricted to $C$. Note that
$\Delta_C = T_C - 2d P_C$ is the difference of the two commuting matrices
$T_C$ and $2d P_C$. Hence, for $x \in C$,
\begin{equation} \label{eq-IIc.30,2}
\big( \exp[\beta \, \Delta_C] \big)_{x,x} 
\ = \
\big( \exp[\beta \, T_C] \: \exp[- 2d \, \beta \, P_C]  \big)_{x,x} 
\ = \
e^{-2d\beta} \: \big( \exp[\beta \, T_C]  \big)_{x,x} \period
\end{equation}
Due to this identity and the fact that $x \in A \subseteq B$, 
Eq.~(\ref{eq-IIc.30,1}) is equivalent to
\begin{equation} \label{eq-IIc.30,3}
\big( \exp[\beta \, T_A] \big)_{x,x} 
\ \leq \
\big( \exp[\beta \, T_B] \big)_{x,x} \period
\end{equation}
Now, $0 \leq (T_A)_{w,z} \leq (T_B)_{w,z}$, and hence
$(T_A^n)_{x,x} \leq (T_B^n)_{x,x}$, for all intergers $n $.
Thus, (\ref{eq-IIc.30,3}) follows from an expansion of the exponentials
in Taylor series,  
\begin{equation} \label{eq-IIc.30,4}
\big( \exp[\beta \, T_A] \big)_{x,x} 
\ = \ 
\sum_{n=0}^\infty \frac{\beta^n}{n!} \, (T_A^n)_{x,x}
\ \leq \
\sum_{n=0}^\infty \frac{\beta^n}{n!} \, (T_B^n)_{x,x}
\ = \ 
\big( \exp[\beta \, T_B] \big)_{x,x} \period
\end{equation}
\hfill {\bf  QED}

\begin{lemma} \label{lem-IIc.3}
Let $Q = \{-m, \ldots, m\}^d \subset \ZZ^d$ be a cube.
Denote by $\Delta_Q$ the nearest-neighbor Laplacian on $Q $,
i.e., $\Delta_Q = P_Q \Delta P_Q = -2d P_Q + T_Q$, $T_Q := P_Q T P_Q$,
and $T_{x,y} = \one( |x-y|_1 = 1 )$. Then, for all $\beta >0$,
\begin{equation} \label{eq-IIc.40,1}
\frac{1}{|Q|} \sum_{z \in Q} \big( \exp[\beta \Delta_{Q}] \big)_{z,z}
\ \geq \
\frac{e^{-d \beta / m}}{(2\pi)^d} 
\int_{[-\pi,\pi]^d} \exp[-\beta \, \om(k)] \: d^dk \comma
\end{equation}
where $\om(k) := \sum_{\nu=1}^d 2 \big\{ 1 - \cos(k_\nu) \big\}$.
\end{lemma}
\Proof We may pick an even integer $r$, choose $L := r \cdot (2m+1)$,
and identify $Q$ with $Q + L\ZZ^d \subseteq \Lambda$. (Note that the
statement of the lemma makes no reference to the Hubbard model
analyzed before, and for the purpose of the proof, $L$ can be taken an
arbitrarily large integer multiple of $2m+1$.) Given $s \in \ZZ_r^d$,
we define $Q(s) := Q + (2m+1)s$ and observe that the family
$\{Q(s)\}_{s \in \ZZ_r^d}$ of cubes define a disjoint partition of
$\Lambda$, i.e.,
\begin{equation} \label{eq-IIc.40,2}
\Lambda \ = \ \bigcup_{s \in \ZZ_r^d} Q(s) 
\hspace{8mm} \mbox{and} \hspace{8mm} 
\forall s \neq s' : \ \ Q(s) \cap Q(s') = \emptyset \period
\end{equation}
Hence
\begin{equation} \label{eq-IIc.40,3}
\hDelta \ := \ 
\sum_{s \in \ZZ_r^d} \Delta_{Q(s)}
\end{equation}
is the sum of translated, but mutually disconnected copies 
of $\Delta_Q$. We observe that
\begin{eqnarray} \label{eq-IIc.40,4}
\lefteqn{
\Tr\big\{ \exp[\beta \hDelta] \big\}
}
\\ \nonumber 
& = &
\sum_{x \in \Lambda} 
\big( \exp[\beta \hDelta] \big)_{x,x}
\ \; = \ \;
\sum_{s \in \ZZ_r^d} \sum_{z \in Q} 
\big( \exp[\beta \hDelta] \big)_{z + (2m+1)s, z + (2m+1)s}
\\ \nonumber 
& = &
\sum_{s \in \ZZ_r^d} \sum_{z \in Q} 
\big( \exp[\beta \Delta_{Q(s)}] \big)_{z + (2m+1)s, z + (2m+1)s}
\ \; = \ \;
r^d \: \sum_{z \in Q} 
\big( \exp[\beta \Delta_{Q}] \big)_{z,z} \period
\end{eqnarray}
As an intermediate result, we thus have 
\begin{equation} \label{eq-IIc.40,5}
\frac{1}{|Q|} \sum_{z \in Q} \big( \exp[\beta \Delta_{Q}] \big)_{z,z}
\ = \
\frac{1}{|\Lambda|} \, \Tr\big\{ \exp[\beta \hDelta] \big\} \comma 
\end{equation}
since $|\Lambda| = L^d = r^d |Q|$.

Next, we translate $\hDelta$ by the elements of $Q$, i.e., for
$\eta \in Q$, we introduce $\hDelta^{(\eta)}$ on $\CC^\Lambda$ by 
\begin{equation} \label{eq-IIc.40,6}
\hDelta^{(\eta)} \ := \ 
\sum_{q \in \ZZ_r^d} \Delta_{Q(q) + \eta}
\ = \ 
\sum_{q \in \ZZ_r^d} \Delta_{Q + \eta + (2m+1)q} \period
\end{equation}
Of course, $\hDelta^{(\eta)}$ is unitarily equivalent to $\hDelta$. 
We observe that 
\begin{equation} \label{eq-IIc.40,7a}
\frac{1}{|Q|} \sum_{\eta \in Q} \hDelta^{(\eta)}
\ = \ 
\frac{1}{|Q|} \sum_{y \in \Lambda} \Delta_{Q + y}
\ = \
-2d \cdot \one_{\CC^\Lambda} \: + \:  
\frac{1}{|Q|} \sum_{y \in \Lambda} T_{Q + y} \comma
\end{equation}
where, for $w,z \in \Lambda$,  
\begin{eqnarray} \label{eq-IIc.40,7b}
\lefteqn{
\Big( \sum_{y \in \Lambda} T_{Q + y} \Big)_{w,z}
\ \; = \ \; 
\sum_{y \in \Lambda} \one_Q(w-y) \, \one_Q(z-y) \, T_{w,z}
}
\\ \nonumber 
& = & 
\big|(Q+w) \cap (Q+z)\big| \cdot T_{w,z} 
\ \; = \ \;
2m \, (2m+1)^{d-1} \: T_{w,z} \comma
\end{eqnarray}
since $T_{w,z} \neq 0$ only if $w-z$ are neighboring lattice sites.
Hence,
\begin{eqnarray} \label{eq-IIc.40,7c}
\frac{1}{|Q|} \sum_{\eta \in Q} \hDelta^{(\eta)}
& = & 
-2d \cdot \one_{\CC^\Lambda} \; + \; \frac{2m}{2m+1} \, T 
\ \; = \ \; 
-\frac{2d}{2m+1} \cdot \one_{\CC^\Lambda} \; + \; \frac{2m}{2m+1} \, \Delta 
\nonumber \\ 
& \geq & 
-\frac{d}{m} \cdot \one_{\CC^\Lambda} \: + \: \Delta 
\end{eqnarray}
where $\Delta \leq 0$ is the nearest-neighbor Laplacian on $\Lambda$
(with periodic b.c.). This and the convexity of 
$A \mapsto \Tr\{e^{\beta A}\}$ therefore imply that 
\begin{eqnarray} \label{eq-IIc.40,8}
\Tr\big\{ \exp[\beta \hDelta] \big\}
& = & 
\frac{1}{|Q|} \sum_{\eta \in Q} 
\Tr\big\{ \exp[\beta \hDelta^{(\eta)}] \big\}
\ \geq \ 
\Tr\bigg\{ \exp\Big[  
\frac{\beta}{|Q|} \sum_{\eta \in Q} \hDelta^{(\eta)} \Big] \bigg\} 
\nonumber \\ 
& \geq & 
e^{- \beta d/ m} \: \Tr\big\{ \exp[\beta \Delta] \big\} \period
\end{eqnarray}
We diagonalize $\Delta$ by discrete Fourier transformation on $\CC^\Lambda$.
The eigenvalues of $-\Delta$ are given by $\om(k)$, where
$k \in \Lambda^* = \frac{2\pi}{L} \ZZ_L^d$ is the variable dual to 
$x \in \Lambda$. Since $|\Lambda^*| = L^d = |Q| \, r^d$, we therefore have
\begin{equation} \label{eq-IIc.40,9}
\frac{1}{|Q|} \sum_{z \in Q} \exp[\beta \Delta_{Q}]_{z,z}
\ = \ 
\frac{1}{|\Lambda|} 
\Tr\big\{ \exp[\beta \hDelta] \big\} 
\ \geq \ 
\frac{e^{- \beta d/ m}}{|\Lambda^*|} 
\sum_{k \in \Lambda^*} e^{-\beta \, \om(k)} \period
\end{equation}
Inequality~(\ref{eq-IIc.40,9}) holds for every $L = r (2m+1)$, and
hence also in the limit $L \to \infty$. Since the right side of
(\ref{eq-IIc.40,6}) is a Riemann sum approximation to the integral in
(\ref{eq-IIc.40,1}), this limit yields the asserted estimate
(\ref{eq-IIc.40,1}).  \hfill {\bf QED}

\subsection{The Discrete Laplacians on $\Om$, $\Om^c$, and their Eigenvalue Sums} 
\label{subsec-IId}
%
In this final subsection, we compare the sum of the eigenvalues of
\begin{equation} \label{eq-IId.0,1}
-\tDelta 
\ := \ 
P_\Om \, (-\Delta) \, P_\Om \; + \; 
P_\Om^\perp \, (-\Delta) \, P_\Om^\perp
\end{equation}
below $\mu$ to the sum of the eigenvalues of $- \Delta$
below $\mu$, where $\Om \subseteq \Lambda$ is an arbitrary, but
henceforth fixed, subset of $\Lambda$, and $\Om^c := \Lambda \setminus
\Om$ is its complement. To this end, we introduce the difference
of these eigenvalue sums,
\begin{eqnarray} \label{eq-IId.1}
\deltaE(\mu, \Om)
& := & 
\Tr\{ [- \tDelta - \mu ]_- \} \; - \; 
\Tr\{ [- \Delta - \mu ]_- \} 
\\ \nonumber
& = &
\Tr\{ (- \tDelta - \mu ) \, \tP_- \} \; - \; 
\Tr\{ (- \Delta - \mu ) \, P_- \} \comma
\end{eqnarray}
where $\tP_- := \one[ -\tDelta \leq \mu]$ and 
$P_- := \one[ -\Delta \leq \mu]$. We further set $\tP_+ := \tP_-^\perp$ 
and $P_+ := P_-^\perp$. Since $\tP_-$ commutes with $P_\Om$, we
have that $\Tr\{ (- \tDelta - \mu ) \, \tP_- \} =
\Tr\{ (- \Delta - \mu ) \, \tP_- \}$, and thus
\begin{eqnarray} \label{eq-IId.2}
\deltaE(\mu, \Om)
& = & 
\Tr\{ (- \Delta - \mu ) \, (\tP_- - P_-) \} 
\\ \nonumber
& = &
\Tr\{ [- \Delta - \mu ]_- \, (\tP_- - \one) \}  \; + \; 
\Tr\{ [- \Delta - \mu ]_+ \, \tP_- \} 
\\ \nonumber
& = &
\Tr\{ [\Delta + \mu ]_+ \, \tP_+ \}  \; + \; 
\Tr\{ [- \Delta - \mu ]_+ \, \tP_- \} 
\ \; \geq \ \; 0 
\end{eqnarray}
is manifestly nonnegative. The derivation of a nontrivial lower bound on
$\deltaE(\mu, \Om)$ of the form $\deltaE(\mu, \Om) \geq \alpha(\mu) \,
|\pOm|$, where $\alpha(\mu) >0$ is a positive constant which depends
only on $\mu$ and the spatial dimension $d \geq 1$ (but not on $\Om$),
is a  task that was first addressed by Freericks, Lieb,
and Ueltschi in \cite{FreericksLiebUeltschi2002}. Shortly thereafter,
Goldbaum \cite{Goldbaum2003} improved the numerical value for
$\alpha(\mu) >0$, especially if $\mu$ is close to $2d$. As a
consequence of the estimates in
\cite{FreericksLiebUeltschi2002,Goldbaum2003}, we have the following
lemma.
\begin{lemma}[Freericks, Lieb, and Ueltschi (2002), Goldbaum (2003)] 
\label{lem-IId.1} \hspace*{\fill} \\
\textbf{(i)} Let $\frac{1}{2} < \mu < 4d$.
There is $L_*(\mu) < \infty$ and $\alpha(\mu) >0$ such that,
for all $L \geq L_*(\mu)$ and all subsets $\Om \subseteq \Lambda$,
\begin{equation} \label{eq-IId.3}
\deltaE(\mu, \Om) \ \geq \ \alpha(\mu) \; |\pOm| \period
\end{equation}
\textbf{(ii)} Let $0< \mu \leq \frac{1}{2}$, and define
\begin{equation} \label{eq-IId.4}
\alpha(\mu) 
\ := \
\frac{|S^{d-1}| \: \mu^{(2+d)/2}}{2^{1+d/2} \, (2\pi)^d \, (4d)^5} 
\hspace{4mm} \mbox{and} \hspace{4mm} 
L_*(\mu) 
\ := \
\frac{4 \pi d}{\mu} \period
\end{equation}
where $|S^{d-1}|$ is the surface volume of the $d$-dimensional sphere. 
Then, for all $L \geq L_*(\mu)$ and all subsets 
$\Om \subseteq \Lambda = \ZZ_L^d$, we have
\begin{equation} \label{eq-IId.5}
\deltaE(\mu, \Om) \ \geq \ \alpha(\mu) \; |\pOm| \period
\end{equation}
\end{lemma}
\Proof
We only give the proof of (ii), which amounts to reproducing
the proof of Lemma~3.1 in \cite{FreericksLiebUeltschi2002}.
By $\{ \psi_k \}_{k \in \Lambda^*} \subseteq \CC^\Lambda$ we
denote the orthonormal basis (ONB) of eigenvectors of $\Delta$, i.e.,
\begin{equation} \label{eq-IId.6}
\psi_k(x) \ := \ |\Lambda|^{-1/2} \: e^{-i k \cdot x} \comma
\quad 
k \: \in \: \Lambda^* \: = \: \frac{2\pi}{L} \ZZ_L^d \comma
\end{equation}
and we have that $-\Delta \psi_k = \om(k) \psi_k$, with
$\om(k) = \sum_{\nu=1}^d 2 \{1 - \cos(k_\nu)\}$. Evaluating the traces
in Eq.~(\ref{eq-IId.2}) by means of this ONB, we obtain
\begin{eqnarray} \label{eq-IId.7}
\deltaE(\mu, \Om) 
& = &
\sum_{k \in \Lambda^*} \Big\{ 
[\mu - \om(k)]_+ \: \la \psi_k | \; \tP_+ \, \psi_k \ra \; + \;
[\om(k) - \mu ]_+ \: \la \psi_k | \; \tP_- \, \psi_k \ra \Big\} \period
\nonumber \\
& \geq & 
\sum_{k \in \Lambda^*} 
[\mu - \om(k)]_+ \: \la \psi_k | \; \tP_+ \, \psi_k \ra  \period
\end{eqnarray}
Let $\{ \vphi_j \}_{j = 1}^{|\Lambda|} \subseteq \CC^\Lambda$ be an 
ONB of eigenvectors of $\tDelta$, i.e., $-\tDelta \vphi_j = e_j \vphi_j$.
For any $k \in \Lambda^*$ and $1 \leq j \leq |\Lambda|$, we observe that
\begin{eqnarray} \label{eq-IId.8}
\big(e_j - \om(k) \big)^2 |\la \psi_k | \vphi_j \ra|^2
& = &
|\la \psi_k | (\Delta - \tDelta) \vphi_j \ra|^2 
\nonumber \\[1mm]
& = &
|\la \psi_k | 
(P_\Om \Delta P_\Om^\perp + P_\Om^\perp \Delta P_\Om) \vphi_j \ra |^2
\nonumber \\[1mm]
& = &
| \la P_\Om \Delta P_\Om^\perp \psi_k | \vphi_j \ra |^2
\: + \:
| \la P_\Om^\perp \Delta P_\Om \psi_k | \vphi_j \ra |^2
\nonumber \\[1mm]
& \geq &
|\la P_\pOm \Delta P_\Om^\perp \psi_k | \vphi_j \ra |^2 \comma
\end{eqnarray}
using that either $P_\Om \vphi_j = 0$ or $P_\Om^\perp \vphi_j = 0$
and that $P_\Om \Delta P_\Om^\perp = P_\pOm \Delta P_\Om^\perp$.
Since $|e_j - \om(k)| \leq 4d$, Eq.~(\ref{eq-IId.8}) implies that 
\begin{equation} \label{eq-IId.9}
(4d)^2 \, |\la \psi_k | \vphi_j \ra|^2
\ \geq \ 
| \la b_k | \vphi_j \ra|^2 \comma
\end{equation}
where $b_k := P_\pOm \Delta P_\Om^\perp \psi_k$ is the boundary vector
that plays a crucial role in \cite{FreericksLiebUeltschi2002}.
By summation over all $j$ corresponding to eigenvalues $e_j > \mu$,
we obtain
\begin{equation} \label{eq-IId.10}
\la \psi_k | \tP_+ \, \psi_k \ra
\ \geq \ 
(4d)^{-2} \, 
\la b_k | \tP_+ \, b_k \ra \comma
\end{equation}
for all $k \in \Lambda^*$. Next, the convexity of 
$\lambda \mapsto [\lambda]_+$ and the fact that 
$\tP_+ = \one[-\tDelta > \mu] \geq (4d)^{-1} [-\tDelta - \mu]_+$
yield
\begin{eqnarray} \label{eq-IId.11}
\la b_k | \tP_+ \, b_k \ra
& \geq & 
\frac{1}{4d} \, \la b_k | [-\tDelta - \mu]_+ \: b_k \ra
\ \; \geq \ \; 
\frac{1}{4d} \, \big[ \la b_k | (-\tDelta - \mu) \: b_k \ra \big]_+ 
\nonumber \\
& = &
\frac{1}{4d} \, \big[ \la b_k | 
\, (-\Delta - \mu) \, b_k \ra \big]_+ \period
\end{eqnarray}
Now, for any $x \in \pOm$ there is, by definition, at least one point 
$x+e \in \Om^c$, with $|e|_1 = 1$. Since $b_k$ is supported in $\pOm$, 
we have $b_k(x+e)=0$, and thus
\begin{eqnarray} \label{eq-IId.12}
\la b_k | (-\Delta-\mu) \, b_k \ra
& = & 
\sum_{x \in \pOm} 
\Big\{ \sum_{|e|_1=1} |b_k(x) - b_k(x+e)|^2 \; - \; \mu |b_k(x)|^2 \Big\}
\nonumber \\
& \geq & 
(1-\mu) \, \sum_{x \in \pOm} |b_k(x)|^2
\ = \ (1-\mu) \, \|b_k\|^2 \period
\end{eqnarray}
Inserting (\ref{eq-IId.10})--(\ref{eq-IId.12}) into (\ref{eq-IId.7}),
we arrive at
\begin{equation} \label{eq-IId.13}
\deltaE(\mu, \Om) 
\ \geq \
\frac{(1-\mu)}{(4d)^3} 
\sum_{k \in \Lambda^*} [\mu - \om(k)]_+ \: \| b_k \|^2 \period
\end{equation}
Next, we use that in the sum in (\ref{eq-IId.13}) only those
$k \in \Lambda^*$ contribute, for which 
$\om(k) = \sum_{\nu =1}^d 2 \big\{1 -\cos(k_\nu) \big\} \leq \frac{1}{2}$,
as $0 < \mu \leq 1$. This implies that 
$\cos(k_\nu) \geq \frac{1}{2}$, for all $\nu \in \{1,2,\ldots,d\}$. 
Hence, for these $k$, we have that
\begin{eqnarray} \label{eq-IId.14}
\| b_k \|^2
& = &
\frac{1}{|\Lambda|} \sum_{x \in \pOm}
\Big| \sum_{\sigma = \pm} \sum_{\nu=1}^d e^{i \sigma k_\nu} \: 
\one[ x + \sigma e_\nu \in \Om^c] \Big|^2
\nonumber \\
& \geq &
\frac{1}{|\Lambda|} \sum_{x \in \pOm}
\Big( \sum_{\sigma = \pm} \sum_{\nu=1}^d \cos(k_\nu) \: 
\one[ x + \sigma e_\nu \in \Om^c] \Big)^2
\nonumber \\
& \geq &
\frac{1}{4 |\Lambda|} \sum_{x \in \pOm} 1
\ \; = \ \;
\frac{|\pOm|}{4 |\Lambda|} \comma
\end{eqnarray}
since there is at least one choice for $(\sigma,\nu)$ such that
$x + \sigma e_\nu \in \Om^c$. Inserting this estimate into
(\ref{eq-IId.13}), we obtain
\begin{equation} \label{eq-IId.15}
\deltaE(\mu, \Om) 
\ \geq \
\frac{|\pOm|}{8 \, (4d)^3} \Big( \frac{1}{|\Lambda^*|} 
\sum_{k \in \Lambda^*} [\mu - \om(k)]_+ \Big) \period
\end{equation}
Now define $q: \TT^d \to \Lambda^*$ by the preimages
\begin{equation} \label{eq-IId.16}
q^{-1}(k) 
\ := \ 
k + \Big[ -\frac{\pi}{L} \, , \, \frac{\pi}{L} \Big)^d \comma
\end{equation}
for $k \in \Lambda^*$. In other words, given $\xi \in \TT^d$, the point 
$q(\xi) \in \Lambda^*$ is the closest point to $\xi$. In particular,
$|\xi - q(\xi)|_{\infty} \leq \frac{\pi}{L}$, which implies that
$|\om(q(\xi)) - \om(\xi)| \leq \frac{2\pi d}{L}$, by Taylor's theorem.
Hence,
\begin{eqnarray} \label{eq-IId.17}
\frac{1}{|\Lambda^*|} 
\sum_{k \in \Lambda^*} [\mu - \om(k)]_+ 
& = &
\int_{\TT^d}  \big[ \mu - \om(q(\xi)) \big]_+ \: \frac{d^d\xi}{(2\pi)^d} 
\nonumber \\
& \geq &
\int_{\TT^d}  \big[ \mu - 2\pi d L^{-1} - \om(\xi) \big]_+ \: 
\frac{d^d\xi}{(2\pi)^d} \period
\end{eqnarray}
Since, by assumption, 
$\frac{2\pi d}{L} \leq \frac{2\pi d}{L_*} = \frac{\mu}{2}$ 
and $\om(\xi) \leq \xi^2$, we have
\begin{equation} \label{eq-IId.18}
\int_{\TT^d}  \big[ \mu - 2\pi d L^{-1} - \om(\xi) \big]_+ \: d^d\xi 
\ \geq \
\int_{\TT^d}  \Big[ \frac{\mu}{2} - \xi^2 \Big]_+ \: d^d\xi 
\ = \
\frac{|S^{d-1}|}{2^{d/2} \, d(d+2)} \: \mu^{1+(d/2)} \period
\end{equation}
Inserting (\ref{eq-IId.17})--(\ref{eq-IId.18}) into (\ref{eq-IId.15}),
we arrive at the asserted estimate.
\hfill {\bf  QED}


\end{document}